\documentclass{aa}  
\usepackage{graphicx}
\usepackage{txfonts}
\usepackage{natbib}
\bibpunct{(}{)}{;}{a}{}{,}
\usepackage{enumitem}
\usepackage[utf8]{inputenc}
\usepackage{epstopdf}
\usepackage[percent]{overpic}
\usepackage{color}
\usepackage{url}
\usepackage[]{xcolor}
\usepackage{tabularx,colortbl}
\usepackage{multirow}
\usepackage{comment}
\newcommand{\ca}{\mbox{Ca\,{\sc ii}~K\,}}

\begin{document}
\authorrunning{Chatzistergos et al.}
\titlerunning{Analysis of full disc H$\alpha$ observations}
\title{Analysis of full disc H$\alpha$ observations: Carrington maps and filament properties over 1909--2022}
\author{Theodosios~Chatzistergos\inst{1}, 
	Ilaria~Ermolli\inst{2}, 
	Dipankar~Banerjee\inst{3},
	Teresa~Barata\inst{4,5},
	Ioannis~Chouinavas\inst{6}, 
	Mariachiara~Falco\inst{7},  
	Ricardo~Gafeira\inst{4,8},
	Fabrizio~Giorgi\inst{2},   
	Yoichiro Hanaoka\inst{9},
	Natalie~A.~Krivova\inst{1},  	
	Viktor V. Korokhin\inst{10}, 
	Ana~Lourenço\inst{4},       
	Gennady P. Marchenko\inst{10}, 
	Jean-Marie~Malherbe\inst{11,12},
	Nuno~Peixinho\inst{4,8},
	Paolo~Romano\inst{7},
	Takashi Sakurai\inst{9}
	}
\offprints{Theodosios Chatzistergos  \email{chatzistergos@mps.mpg.de}}
\institute{Max Planck Institute for Solar System Research, Justus-von-Liebig-weg 3,	37077 G\"{o}ttingen, Germany 
	\and INAF Osservatorio Astronomico di Roma, Via Frascati 33, 00078 Monte Porzio Catone, Italy 
	\and Aryabhatta Research Institute of Observational Sciences, Manora peak, Nainital-263 001, India 
	\and Instituto de Astrofísica e Ciências do Espaço, University of Coimbra, 3034-004 Coimbra, Portugal
	\and Department of Earth Sciences, University of Coimbra, 3030-790 Coimbra, Portugal
	\and Larissa Observatory 'Aristeus', 41500 Larissa, Greece
	\and INAF Osservatorio Astrofisico di Catania, Catania, Italy
	\and Department of Physics, University of Coimbra, 3004-516 Coimbra
	\and National Astronomical Observatory of Japan 2-21-1 Osawa, Mitaka, Tokyo 181-8588, Japan
	\and Astronomical Institute of Kharkiv V.N. Karazin National University, 35 Sumskaya St., Kharkiv 61022, Ukraine
	\and LESIA, Observatoire de Paris, 92195 Meudon, France
    \and PSL Research University, Paris, France
  }
\date{}
	
\abstract
{Full disc observations of the Sun in the H$\alpha$ line provide information about the solar chromosphere and in particular about the filaments, which are dark and elongated features that lie along magnetic field polarity inversion lines.
This makes them important for studies of solar magnetism.
Since full disc H$\alpha$ observations have been performed at various sites since the second half of the 19th century, with regular photographic data having started in the beginning of the 20th century, they are an invaluable source of information 
on past solar magnetism. }
{In this work we aimed at deriving accurate information about filaments from historical and modern full disc H$\alpha$ observations.} 
{We have consistently processed observations from 15 H$\alpha$ archives spanning 1909--2022. The analysed datasets include long-running ones such as those from Meudon and Kodaikanal, but also previously unexplored datasets such as those from Arcetri, Boulder, Larissa, and Upice. 
Our data processing includes photometric calibration of the data stored on photographic plates, the compensation of the limb-darkening and orientation of the data to align the solar north at the top of the images. 
We have constructed also Carrington maps from the calibrated H$\alpha$ images.}
{We find that filament areas, similar to plage areas in \ca data, are affected by the bandwidth of the observation.
Thus, cross-calibration of the filament areas derived from different archives is needed.
We have produced a composite of filament areas from individual archives by scaling all of them to the Meudon series.
Our composite butterfly diagram shows very distinctly the common features of filament evolution, that is the poleward migration as well as a decrease in the mean latitude of filaments as the cycle progresses.
We also find that during activity maxima, filaments on average cover $\sim$1\% of the solar surface.
We see only a weak change in the amplitude of cycles in filament areas, in contrast to sunspot and plage areas.} 
{Analysis of H$\alpha$ data for archives with contemporaneous \ca observations allowed us to identify and verify archive inconsistencies, which will also have implications for reconstructions of past solar magnetism and irradiance from \ca data.}
	
	
\maketitle
	
\section{Introduction}
\label{sec:intro}
\sloppy
\begin{figure*}[t!]   
	\centering 
		\includegraphics[width=1\linewidth]{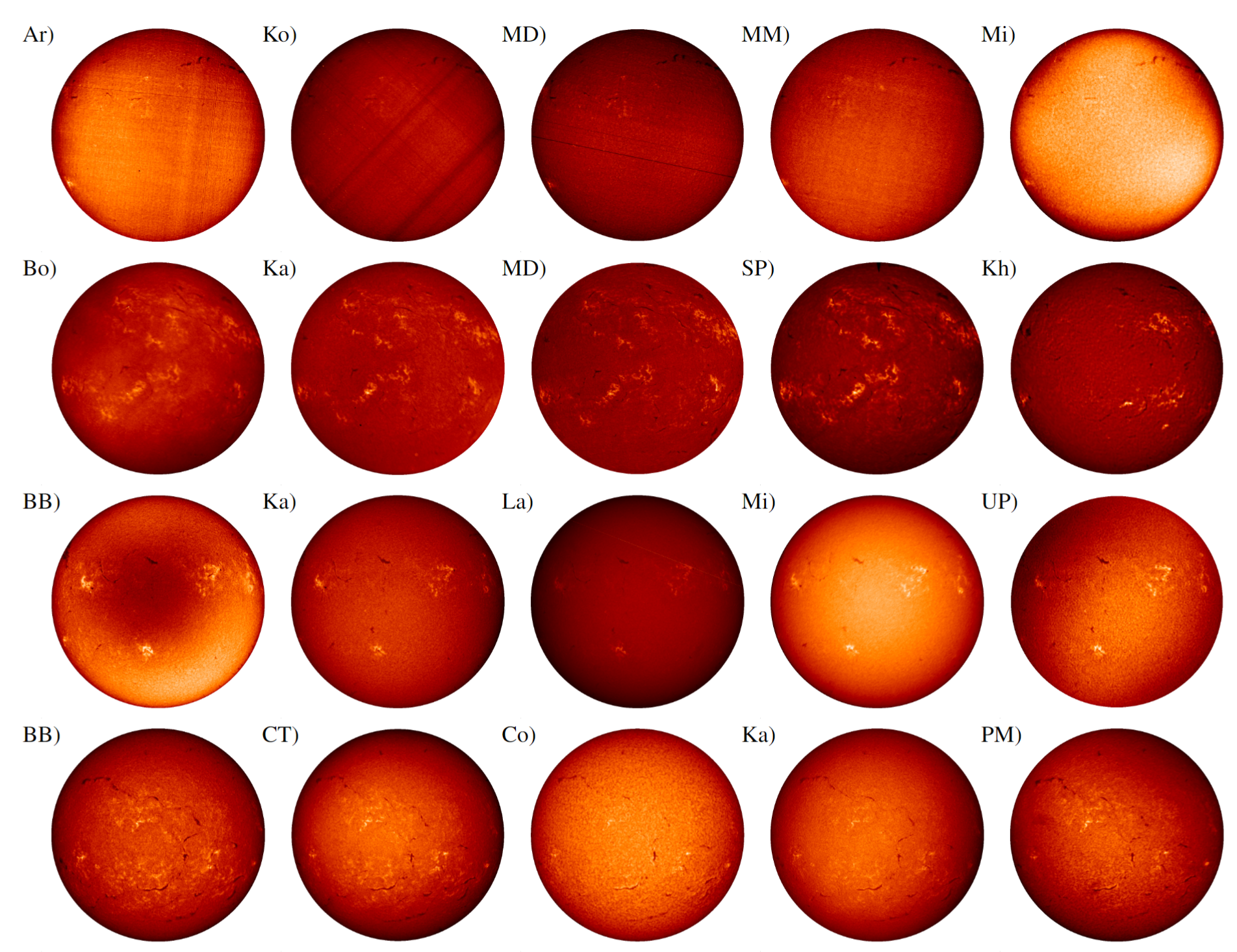}
\caption{Examples of H$\alpha$ observations from the various archives analysed in this study. The images within each row correspond to the same day, with the exception of the Kh image (taken on 26 May 2000). In particular, the dates of the observations are: 04 August 1965 for Ar, Ko, MD, MM, and Mi; 14 January 1980 for Bo, Ka, MD, and SP; 02 September 2011 for BB, Ka, La, Mi, and UP; 13 September 2013 for BB, CT, Co, Ka, and PM, respectively. The images are shown after the pre-processing to identify the disc and re-sample them to account for the disc ellipticity (when applicable) and convert the historical data to density values. The images have been aligned to show the solar north pole at the top.}
\label{fig:processedimages_raw} 
\end{figure*}

The first line of the hydrogen Balmer series at 6562.79 \AA, known as H$\alpha$ line, is one of the deepest and broadest lines of the visible spectrum of the Sun. 
\citet{janssen_iii_1869} and \citet{lockyer_v_1869} were the first ones to observe on-disc solar features in H$\alpha$.
In the late 1800's Secchi \citep{secchi_sulla_1871} and Tacchini \citep{tacchini_macchie_1872} made daily observations in the H$\alpha$ line \citep{bocchino_migrazione_1933,ermolli_legacy_2021,ermolli_solar_2023}, and even though these were regular observations they were merely visual inspections and reported in drawings. 
At that time H$\alpha$ observations were not common 
and it was not before the invention of the spectroheliograph \citep[hereafter SHG,][]{hale_development_1904} and daily use of photography that they could be explored more widely and systematically. 
Note that regular photographic observations in the H$\alpha$ line started later than those in white-light and the \ca line because the emulsion of the early photographic plates was not sensitive to wavelengths greater than $\sim$5900~\AA. 
This changed with the invention of red-sensitized emulsions in 1907 \citep{wallace_studies_1907}, which allowed storing H$\alpha$ observations on photographic plates since 1909.
Therefore, H$\alpha$ data make the third longest series of direct photographic full disc observations of the Sun after the white-light observations of the photosphere and \ca images of the chromosphere \citep{chatzistergos_full-disc_2022}. 
Similarly to the \ca line, the H$\alpha$ line has long been used as a tool for extracting information about the structure and dynamics of the solar chromosphere. 

The \ca and H$\alpha$ lines probe slightly different altitudes in the solar atmosphere \citep{vernazza_structure_1981}. 
In fact, the wings and core of the \ca line sample almost continuously the heights from the photosphere to the high chromosphere, while the wings and core of the H$\alpha$ distinctly probe the deep photosphere and the overlying chromosphere, respectively. 
This distinct origin of the \ca  and H$\alpha$ results in a significantly different appearance of the chromosphere. 
While observations in the \ca line, at e.g. 2\AA~spectral resolution, mostly trace bright features associated with photospheric magnetic regions,  H$\alpha$ images show structures mainly tracing the canopy field in the higher atmosphere. 
In fact, when imaged at e.g. 2 \AA~spectral resolution,  \ca observations resemble unsigned magnetograms \citep[][and references therein]{chatzistergos_recovering_2019,murabito_investigating_2023} describing the evolution of the solar surface magnetic field in non-spot regions, while  H$\alpha$ data display a highly structured and dynamic environment filled with fibrils, mottles, filaments, and flaring regions \citep{carlsson_new_2019}.

Fibrils and mottles are short-lived (from a few minutes to a few hours), small-scale (of the order of tens of arcsec) jet-like features that dominate the chromosphere above and about plage regions \citep{de_pontieu_high-resolution_2007} and in quiet Sun, respectively. 
Their counterparts seen at the limb are known as spicules \citep[e.g.][]{tsiropoula_time_1994,rouppe_van_der_voort_-disk_2009,pereira_quantifying_2012}. 

Filaments are long-lived (from days to weeks), large-scale \citep[can reach up to lengths comparable to the solar radius;][]{mazumder_solar_2021} elongated structures of dense and relatively cold gas that appear dark relative to their surroundings in on-disc observations and bright when sticking out beyond the solar limb to form prominences \citep{parenti_solar_2014}. 
These structures, which  are held in place by the magnetic field, lie between areas of opposite polarity magnetic fields \citep{babcock_suns_1955,mcintosh_solar_1972,martin_conditions_1998}.
Prominences have been studied with H$\alpha$ observations \citep[e.g.][with such analyses extending information on prominences back to 1869 with observations by Respighi, Secchi, Tacchini, and Donati from Rome, Palermo, and Florence]{bocchino_migrazione_1933,carrasco_note_2021}, and with disc-blocked or narrow bandwidth \ca observations \citep[e.g.][extending information on prominences back to 1906 with observations from Kodaikanal and Coimbra observatory]{chatterjee_timelatitude_2020,carrasco_catalog_2022}. 
Also filaments can be seen both in H$\alpha$ and \ca observations, however they are less prominent in \ca data and can be seen only in observations acquired  with relatively narrow bandwidths \citep[e.g. as reported in the catalogs of the observatory of Coimbra between 1929--1944;][]{lourenco_testing_2021,carrasco_catalog_2022,wan_differential_2022}. 
Therefore, most of the present knowledge on filaments comes from analysis of observations in the H$\alpha$ line \citep[e.g.][]{hansen_global_1975,makarov_morphology_1982,makarov_poleward_1983-1,makarov_poleward_1983,coffey_digital_1998,mouradian_new_1998,denker_synoptic_1999,shih_automatic_2003,fuller_filament_2005,benkhalil_automated_2005,zharkova_filament_2005,li_activity_2007,yuan_automatic_2011,hao_developing_2013,hao_statistical_2015,zou_non-linear_2014,laurenceau_retrospective_2015,tlatov_tilt_2016-1,chatterjee_long-term_2017,diercke_chromospheric_2019,diercke_solar_2022,lin_new_2020,suo_full-disk_2020}.
It is worth noting that most of the previous studies were based on analysis of observations covering limited time intervals and archives. 
In particular, only the data from Meudon and Kodaikanal observatories have been employed for periods before 1960. 
Moreover, previous studies based on analysis of multiple archives were mainly performed by appending results form the different series without studying potential inconsistencies of the data.

For a long time, the above limitations, which also affected studies using observations at other wavelengths (such as \ca) were essentially imposed by the limited availability of data in digital form and the lack of the needed image processing resources. 
However, over the recent years these limitations have been superseded, allowing a more detailed and careful analysis of existing H$\alpha$ data. 
Indeed, only accurate and consistent analysis of multiple archives would allow accounting for  artefacts in the results derived from the individual datasets and thus a meaningful assessment of underlying solar processes  \citep{ermolli_comparison_2009,ermolli_potential_2018,chatzistergos_analysis_2019,chatzistergos_reconstructing_2021-1}. 
To this goal, \citet{chatzistergos_analysis_2018} introduced a novel approach to process full disc solar observations from historical and modern series available in digital form \citep{chatzistergos_analysis_2017,chatzistergos_exploiting_2016,chatzistergos_ca_2018,chatzistergos_historical_2019,chatzistergos_analysis_2019,chatzistergos_delving_2019,chatzistergos_historical_2020,chatzistergos_analysis_2020}. 
This approach, which includes an accurate photometric calibration of photographic observations and compensation for the centre-to-limb-variation (CLV), was found to return more accurate results than those achieved with other methods in the literature. 
Originally developed to analyse \ca data, this method can be applied to consistently process 
observations from multiple archives and in different spectral bands. 
Indeed, in addition to 43 \ca datasets \citep{chatzistergos_analysis_2020},
it has already been successfully applied to data over four continuum-intervals \citep{chatzistergos_recovering_2019,chatzistergos_modelling_2020,ermolli_rome_2022} and provisionally to a small sample of H$\alpha$ images from the Kanzelhöhe observatory \citep{asvestari_modelling_2021}.

\newcounter{tableid}
\begin{table*}
	\caption{List of H$\alpha$ archives analysed in this study.}
	\label{tab:observatories}
	\centering
	\begin{tabular}{l*{8}{c}}
		\hline\hline
		\textbf{Observatory} & Acronym	   &Detector &Instrument&Period		  &Images&SW   			          &Pixel scale			  	  & Ref.\\
							 &   		   &		 &			& 		      &      &[$\text{\AA}$]		  &[$"/$pixel] 			  	  &	\\
		\hline
		Arcetri				 &Ar		   &Plate	                  & SHG		& 1931--1968  & 5416 	  &0.3   				  &2.3--4.8\tablefootmark{a}   &
		    \addtocounter{tableid}{1}\thetableid, \addtocounter{tableid}{1}\thetableid\\  
		Big Bear			 &BB		   &CCD		                  & Filter	& 1988--2022  & 9527 	  &0.5   				  &4.3, 1.0\tablefootmark{b}   &
		    \addtocounter{tableid}{1}\thetableid\\  
		Boulder				 &Bo		   &Plate	                  & Filter  & 1967--1994  & 6864      & 0.5					  &3.7						   &
		    \addtocounter{tableid}{1}\thetableid\\  
        Catania				 &CT	       &CCD		                  & Filter	& 2002--2022  & 8026 	  &0.5    				  &2.1,1.3,1.1\tablefootmark{c}&
		    \addtocounter{tableid}{1}\thetableid\\  
		Coimbra				 &Co		   &Plate/CCD\tablefootmark{d}& SHG		& 1977--2021  & 8931 	  &0.26   				  &1.0,2.2\tablefootmark{e}    &                                            \addtocounter{tableid}{1}\thetableid\\ 
        Kanzelhöhe			 &Ka		   &Plate/CCD\tablefootmark{f}& Filter	& 1973--2022  & 37276     &0.7   				  &3.0, 2.1, 2.3, 1.0\tablefootmark{g}
            &\addtocounter{tableid}{1}\thetableid, \addtocounter{tableid}{1}\thetableid\\  
		Kharkiv			     &Kh		   &Plate/CCD\tablefootmark{h}                      & SHG 	& 1938--2005  & 256       &0.5   				  &1.8, 4.0\tablefootmark{i}
            &\addtocounter{tableid}{1}\thetableid\\ 
		Kodaikanal			 &Ko		   &Plate	                  & SHG		& 1912--2007  & 30506     &0.5   				  &0.89	  			                                                        &\addtocounter{tableid}{1}\thetableid\\ 
        Larissa              &La           &CCD                       &Filter   &2008--2015   & 100       &0.5                   &1.5, 0.8\tablefootmark{j}&\addtocounter{tableid}{1}\thetableid\\
		McMath-Hulbert		 &MM		   &Plate	                  & SHG		& 1948--1966  & 3728 	  &0.5   				  &3.0	  			                                                        &\addtocounter{tableid}{1}\thetableid\\  
		Meudon				 &MD		   &Plate/CCD\tablefootmark{k}& SHG		& 1909--2022  & 26069 	  &0.25, 0.15\tablefootmark{l}&0.7--4.8, 1.56, 1.1\tablefootmark{m}			          &\addtocounter{tableid}{1}\thetableid\\  
		Mitaka				 &Mi		   &Plate/CCD\tablefootmark{n}& Filter		& 1957--2022  & 34836 	  &0.75, 0.5, 0.25\tablefootmark{o}  				  &2.8, 4.5, 2.8, 1.1\tablefootmark{p}	  			          &\addtocounter{tableid}{1}\thetableid\\  
		Pic du Midi		     &PM		   &CCD	                      & Filter	& 2011--2022  & 3149      &0.5					  &1.2	  			          &\addtocounter{tableid}{1}\thetableid\\  
		Sacramento Peak		 &SP		   &Plate	                  & SHG		& 1960--2002  & 5477      &0.5					  &1.2	  			          &\addtocounter{tableid}{1}\thetableid\\  
		Upice		         &UP		   &CCD	                      & Filter	& 2002--2022  & 1912      &1.6					  &4, 2.5\tablefootmark{q}	  			          &\addtocounter{tableid}{1}\thetableid\\  
		\hline
	\end{tabular}
	\tablefoot{Columns are: name of the observatory, abbreviation, type of detector, type of instrument, period of observations, number of images, spectral width of the spectrograph/filter, pixel scale of the image, and the bibliography entry.
	\tablefoottext{a}{Some images from 1940 and 1931 have pixel scale of $\sim4.8"$ and 
	$\sim3.3"$, respectively while all other observations have a pixel scale of $\sim2.5"$.}
	\tablefoottext{b}{The two values correspond to the periods before and after 27/11/1995.}
	\tablefoottext{c}{The values correspond to the periods [23/07/2002--25/07/2006], [26/07/2006--20/09/2012], and since 21/09/2012, respectively.}
    \tablefoottext{d}{The observations were stored on photographic plates up to 13/01/2007, while observations with a CCD camera started on 01/07/2008.}
    \tablefoottext{e}{The two values correspond to the periods before and after 13/01/2007.}
    \tablefoottext{f}{The observations were stored on photographic plates up to 10/03/2000, while observations with a CCD camera started on 24/09/1998.}
	\tablefoottext{g}{The values correspond to the periods [05/05/1973--31/05/1975], [01/06/1975--10/03/2000], [24/09/1998--22/09/2010], and since 01/06/2008, respectively.}
	\tablefoottext{h}{The observations since 1994 were performed with a CCD camera, while the earlier observations were stored on photographic plates.}
	\tablefoottext{i}{The two values correspond to the data stored on photographic plates and those taken with a CCD camera.}
    \tablefoottext{j}{The two values correspond to the periods before and after 23/09/2008.}
    \tablefoottext{k}{The observations were stored on photographic plates up to 27/09/2002, while observations with a CCD camera started on 13/05/2002.}
	\tablefoottext{l}{The values refer to the periods [01/01/1909--14/06/2017] and [14/06/2017--31/12/2020].}
	\tablefoottext{m}{The values refer to the periods [01/01/1909--27/09/2002], [13/05/2002--14/06/2017], and [14/06/2017--31/12/2020].}
    \tablefoottext{n}{The observations were stored on photographic plates up to 23/04/1992, while observations with a CCD camera started on 28/09/1991.}
	\tablefoottext{o}{0.75\AA~was used for the data from the Monochromatic Heliograph before 20 October 1971 and since 29 March 1974. 0.5\AA~was used for the data from the Monochromatic Heliograph between 20 October 1971 and 28 March 1974 as well as those from the Automatic Flare Patrol Telescope (1991--2019, though in this work we used the data only up to 05/05/2011). 0.25\AA~was used for the Solar Flare Telescope since 06/06/2011.}
	\tablefoottext{p}{The values correspond to the periods [01/10/1957--23/04/1992], [28/09/1991--28/01/2003], [25/09/2008--05/06/2011], and since 06/06/2011 respectively.}
	\tablefoottext{q}{The two values correspond to the period before and after 01/01/2018, when the CCD camera was upgraded.}
	}
	\tablebib{\addtocounter{tableid}{-\thetableid}
		(\addtocounter{tableid}{1}\thetableid) \citet{ermolli_digitized_2009}; 	
		(\addtocounter{tableid}{1}\thetableid) \citet{righini_riduzione_1950};  
		(\addtocounter{tableid}{1}\thetableid) \citet{denker_synoptic_1999};    
		(\addtocounter{tableid}{1}\thetableid) \citet{wu_partial_1975};         
		(\addtocounter{tableid}{1}\thetableid) \citet{romano_evolution_2022};   
		(\addtocounter{tableid}{1}\thetableid) \citet{lourenco_solar_2019};     
		(\addtocounter{tableid}{1}\thetableid) \citet{potzi_scanning_2008};     
		(\addtocounter{tableid}{1}\thetableid) \citet{potzi_kanzelhohe_2021};   
		(\addtocounter{tableid}{1}\thetableid) \citet{belkina_ccd_1996};     
		(\addtocounter{tableid}{1}\thetableid) \citet{chatterjee_long-term_2017}; 
		(\addtocounter{tableid}{1}\thetableid) \url{https://www.larissa.gov.gr/dsites/larobs/SUN.htm}; 
		(\addtocounter{tableid}{1}\thetableid) \citet{mohler_mcmath-hulbert_1968}; 
		(\addtocounter{tableid}{1}\thetableid) \citet{malherbe_new_2019}; 
		(\addtocounter{tableid}{1}\thetableid) \citet{hanaoka_synoptic_2020}; 
		(\addtocounter{tableid}{1}\thetableid) \citet{koechlin_solar_2019}; 
		(\addtocounter{tableid}{1}\thetableid) \citet{tlatov_new_2009}; 
		(\addtocounter{tableid}{1}\thetableid) \citet{klimes_simultaneous_1999}.} 
\end{table*}
\begin{figure*}
	\centering
	\includegraphics[width=1\linewidth]{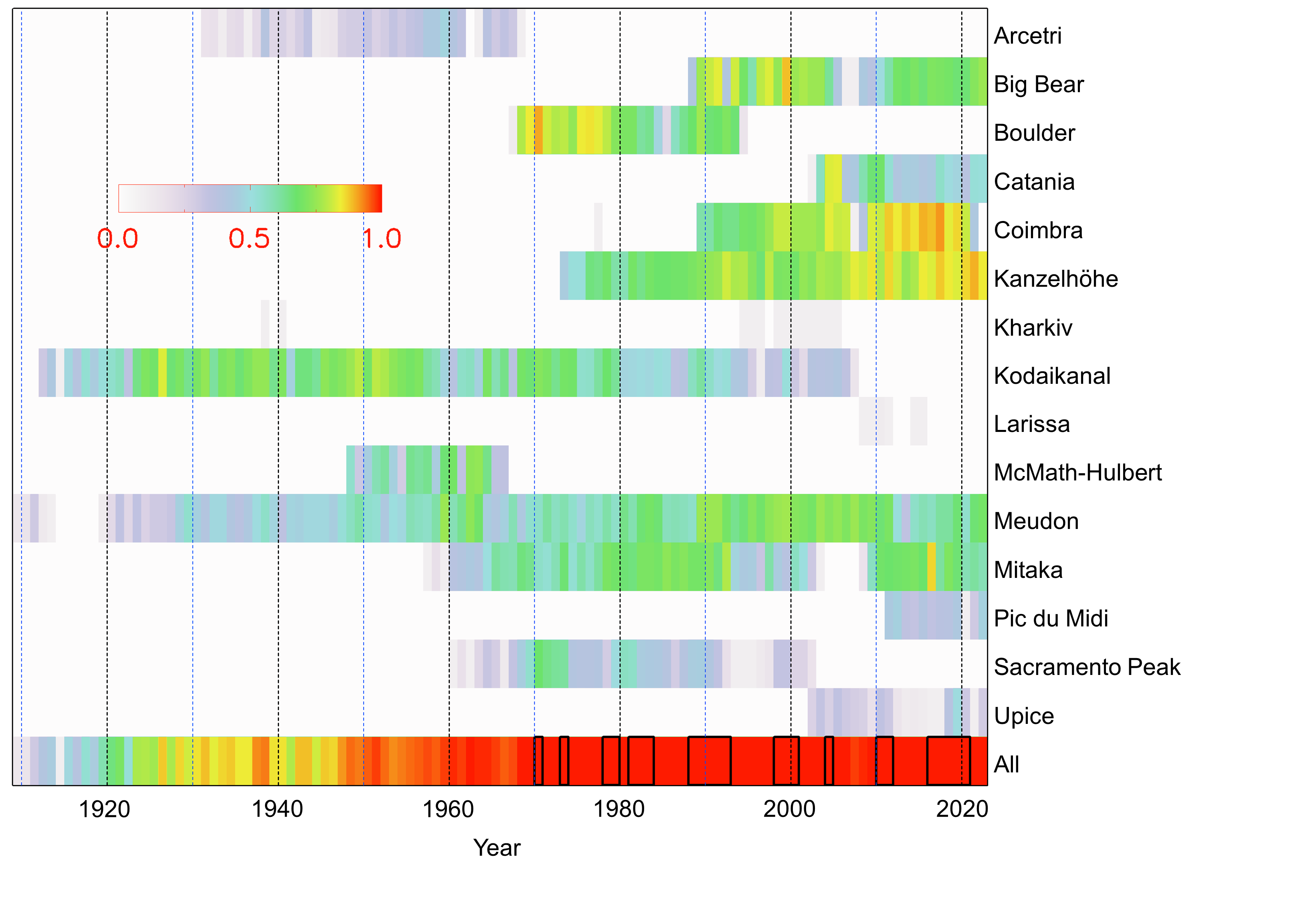}
	\caption{Annual fractional coverage by the various H$\alpha$ archives analysed in this study. Also shown is the annual coverage by all the archives combined. The annual coverage is colour-coded as shown by the colour bar plotted within the plot. Black boxes mark the years with complete daily coverage.}
	\label{fig:timeline}
\end{figure*}

\begin{figure*}[t!]   
	\centering 
		\includegraphics[width=1\linewidth]{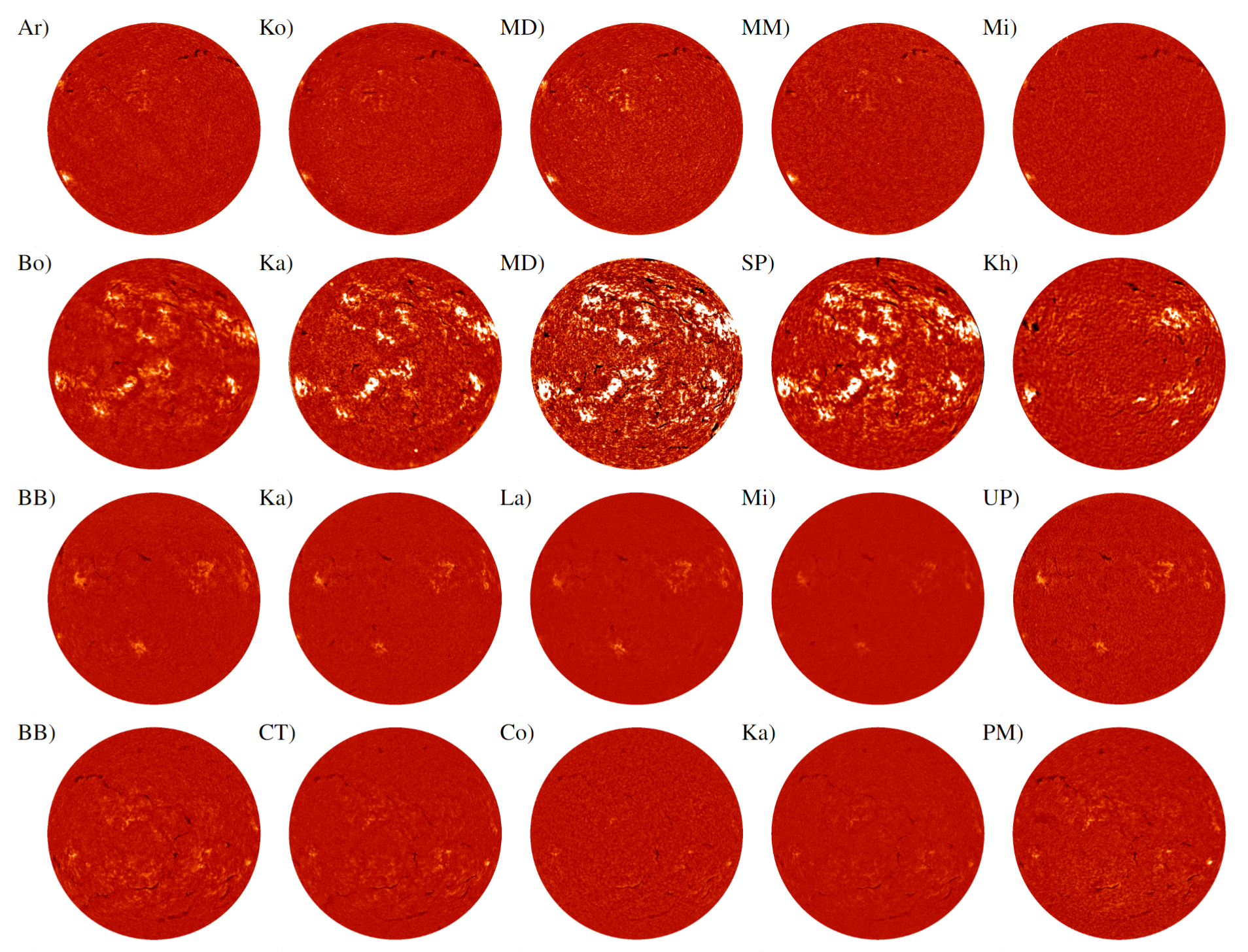}
\caption{Examples of photometrically calibrated and limb-darkening compensated contrast images for the observations shown in Fig. \ref{fig:processedimages_raw}. All images show contrast values in the range [-0.5,0.5].}
\label{fig:processedimages_flat} 
\end{figure*}

\begin{figure*}[t!]   
	\centering 
		\includegraphics[width=1\linewidth]{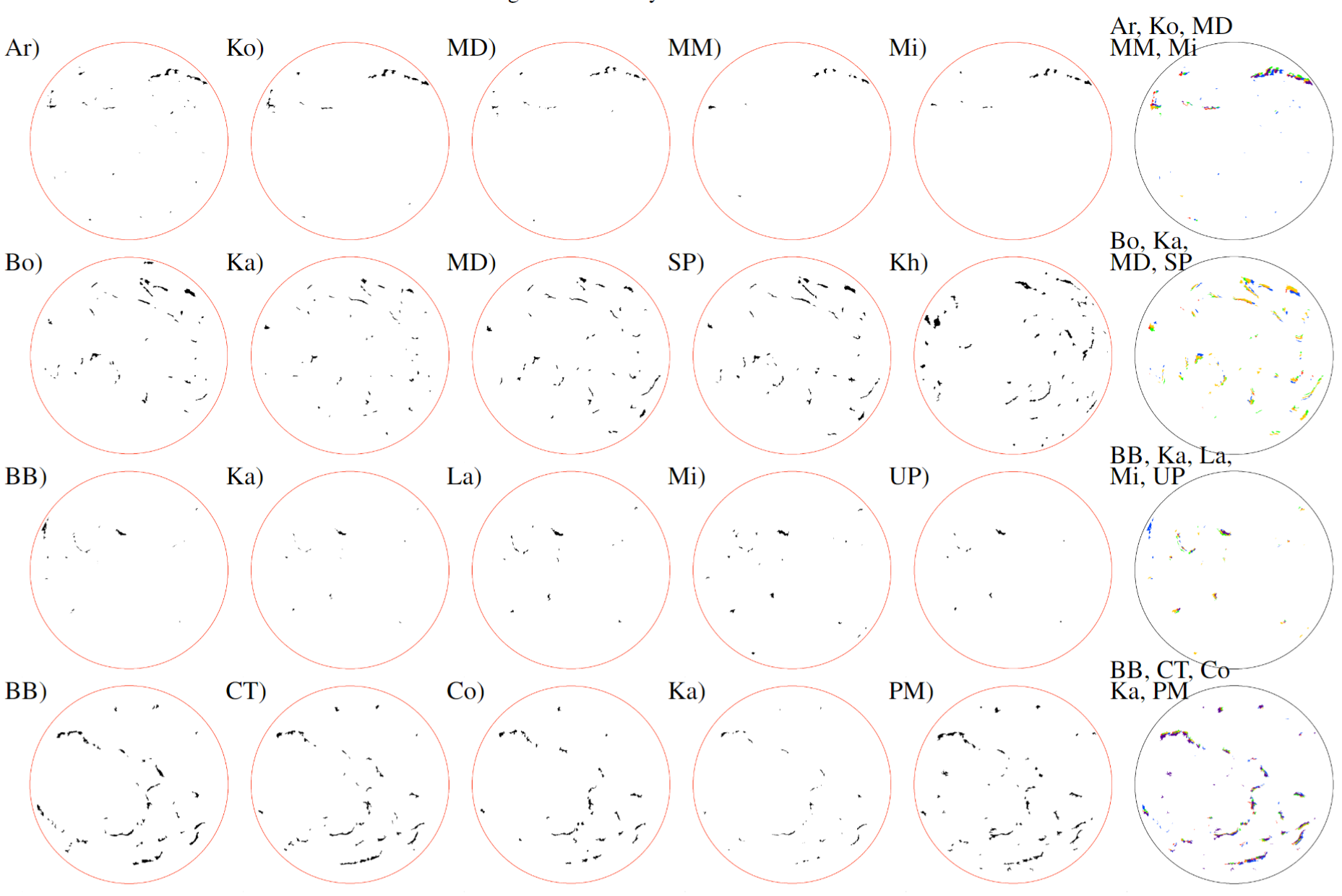}
		\caption{Examples of masks identifying  filaments in the observations shown in Fig. \ref{fig:processedimages_raw}. The circles mark the disc boundaries, while  filaments are shown in black. The last column overlays the masks from the various observations taken on the same day and shown in the corresponding row, coloured blue, red, green, yellow, and purple. We remind that observations in each row are taken on the same day, with the exception of Kh in the second row, which is why it is not overlaid in the last panel of the row.
		On the masks shown in the last column, we have compensated for the differential rotation  so to show all of them as they would have been at 12:00 UTC.}
\label{fig:processedimages_mask} 
\end{figure*}

Here, we analyse H$\alpha$ data 
from 15 digitised archives to produce an extensive and consistent database of filaments covering the period 1909--2022. 
To our knowledge, this study substantially exceeds earlier such efforts. 
Furthermore, nine of the archives analysed in this study have not been used for such studies before, while the pre-1980 data from Meudon have previously only been analysed manually. 
Using multiple archives, we are able to increase the daily coverage of observations available for the analysis of filament evolution and, yet more importantly, to single out trends in the results which are not due to solar processes, but are rather introduced by changes in the instruments and characteristics of individual archives.
Furthermore, for 12 of the analysed archives there are near-co-temporal \ca observations from the same site. 
Thus the present analysis of the H$\alpha$ observations can also improve our understanding of the observational conditions for the \ca observations and vice versa.

The paper is structured as follows.
The data and processing techniques are described in Sect. \ref{sec:datamethods}.
We present the Carrington maps from the H$\alpha$ observations and our results on the filament identification in Sect. \ref{sec:results}.
We summarise and draw our conclusions in Sect. \ref{sec:conclusions}.
In Appendix \ref{sec:archive_characteristics} we compare the characteristics of the various archives.

\section{Data and methods}
\label{sec:datamethods}
\subsection{H$\alpha$ observations}
\label{sec:data}
In this study we used full disc H$\alpha$ (6562.79 \AA) observations from 15 archives.
Table \ref{tab:observatories} lists the main characteristics of these archives.
To our knowledge, the analysed archives comprise all historical datasets that are currently available in digital form  as well as some CCD-based ones that were available to us. 
In particular, we used the data from the Arcetri (Ar), Big Bear (BB), Boulder (Bo), Catania (CT), Coimbra (Co), Kanzelhöhe (Ka), Kharkiv (Kh), Kodaikanal (Ko), Larissa (La), McMath-Hulbert (MM), Meudon (MD), Mitaka (Mi), Pic du Midi (PM), Sacramento Peak (SP), and Upice (UP) observatories.
Figure \ref{fig:processedimages_raw} shows exemplary images from all the analysed archives. 
In total all these archives contain 182,073 images covering 36,991 days over the period 1909--2022.
The temporal coverage is 89\% over the entire period (which is between 19 April 1909, the first day with an H$\alpha$ observation among our analysed archives, and 31 December 2022). The coverage is higher and rather stable at 99.7\% since 1968, while it is steadily decreasing when going further back in time before 1968.
Figure \ref{fig:timeline} gives the annual coverage from each archive separately as well as when all archives are combined together.

The archives analysed in this study include observations taken with diverse settings, while various instrumental changes potentially introduce inconsistencies in the series \citep[similarly to what is reported for \ca data;][]{chatzistergos_full-disc_2022}. 
They comprise observations performed with a spectroheliograph (Ar, Kh, Ko, MM, MD, SP, Co) and optical filters (Mi, Bo, Ka, BB, CT, PM, La, UP). 
The images were stored on photographic plates/film (Ar, Ko, MM, Bo, SP) or taken with a CCD camera (BB, CT, PM, La, UP), while the archives from Mi, Ka, Kh, MD, Co include both CCD-based observations and stored on photographic plates. 
Only a small sample of the Kh collection exists in digital form, which was scanned for the purposes of this study. 
Moreover, the Mi data used in this study derive from three different telescopes, that is the Monochromatic Heliograph (1957--1992), the Automatic Flare Patrol Telescope (1991--2019), and the Solar Flare Telescope (since 2011). 
The data with the latter two telescopes were taken with a CCD camera, while those with the monochromatic heliograph were stored on films which were digitised in the late 1990s to early 2000s with 8-bit depth.
Unfortunately, the films deteriorated and were eventually discarded, but luckily they were properly digitised first.
Since the filter used for the the Automatic Flare Patrol Telescope severely degraded over the later years, we have ignored the data taken with this filter since 2011. 
For this period we used only the Mi data from the solar flare telescope.
Note that MD is the oldest H$\alpha$ archive \citep{malherbe_new_2019,malherbe_130_2023}, with the earliest observations dating back to 1909 and the full series currently covers 10 solar cycles, while the one from La is the shortest series analysed in this study, covering less than one full solar cycle. 

The datasets differ also in terms of cadence of observations.
Most currently running sites (BB, CT, Ka) as well as Bo and Mi have high cadence observations, while all other sites performed only a few observations per day.
For our current analysis we considered 1--3 images per day from the archives with high cadence observations. 
An exception was for Mi for which we analysed all the data taken with the Automatic Flare Patrol Telescope (and a CCD camera) and approximately half of the available data taken with the Monochromatic Heliograph (and stored on photographic plates).
This is because the Monochromatic Heliograph data were scanned in a way that most files include 2 solar observations, typically taken a few seconds apart.
However, we still analysed 1--3 images per day from the Mi data taken with the Solar Flare Telescope.
For the latter we note that the archive includes data with three different exposures, of which we used the medium one so to increase the signal to noise ratio.
An exception was made for observations for which the medium exposure led to saturated plage regions, for which dates we used the low exposure images.
Another issue important for this study is the change from photographic plates to film in the Ko dataset in 1978 \citep{chatterjee_long-term_2017}. 
Unfortunately, most Ko photographs since 16 December 1998 
appear torn, resulting in parts of the solar disc being missing in the observations. 
Notwithstanding this issue, we processed those data, too, and still used them since they can provide information about filaments and plage regions away from the poles. 
We also corrected an inconsistency in the recording of the observational time of Ko data between 01 September 1942 and 15 October 1945 during which period daylight savings time was used \citep{jha_long-term_2022}.

Note that La data have not been described in the literature before.
They were performed at the Larissa observatory\footnote{\url{https://www.larissa.gov.gr/el/i-poli/index.php?option=com_k2&view=item&id=347:asteroskopeio-aristefs-english}} in central Greece, which was founded by Nick Stoikidis in 1972.
The telescope used for these observations has the following characteristics: diameter of objective lens D = 160mm, focal length F = 2400 mm, and focal ratio f/15.
A CCD camera (PRO LINE PL11002M-2 by Finger Lakes Instrumentation) was used producing images with dimensions of 4008 x 2672 pixel$^2$.
There are also more than 2000 35mm film observations over 1991--1993, which however have not yet been digitised.

\subsection{Methods}
All images were consistently processed with the methods developed by \cite{chatzistergos_analysis_2018,chatzistergos_analysis_2019,chatzistergos_analysis_2020}.
Only the process to perform the photometric calibration of the photographic data was adapted to the specifics of the H$\alpha$ line, due to the difference in the quiet Sun (QS) centre-to-limb variation (CLV) in \ca and H$\alpha$ observations. 
In the following we briefly describe the applied processing. 

As a first step, we identified the solar limb with a Sobel filtering followed by an ellipse fit to compute the coordinates of the disc centre and semi-major/minor axes \citep[for more details see Appendix B of][and \citealt{chatzistergos_historical_2020}]{chatzistergos_analysis_2019}. 
This allowed us to account for the recorded solar disc ellipticity by re-sampling the images accordingly \citep{chatzistergos_historical_2020,chatzistergos_analysis_2020}.
The background of the images, comprising the limb darkening and various small- and large-scale artefacts, was computed by applying a running-window median filter along with polynomial fitting across linear and radial segments \citep{chatzistergos_analysis_2018}. 
This processing was applied iteratively such that the solar disc features were progressively excised to improve the accuracy of the background computation and eliminate the influence of bright and dark features in that computation.
The relation between the QS CLV measured on the historical data and a reference QS CLV from CCD-based data allowed us to perform the photometric calibration for each image (only those stored in photographic plates) separately \citep[see][for more details]{chatzistergos_analysis_2018}. 
As the reference QS CLV, we took 
that derived from Catania (CT) CCD-based observations over the period 2002--2018. 
The choice of the CT data as the reference was due to the sufficiently long interval of observations at that site, high consistency of the data, large number of daily observations per year, and the fact that this archive has a nominal bandwidth which is representative of most available H$\alpha$ archives analysed here. 
We produced photometrically calibrated and limb-darkening-compensated contrast images with values given in the form of $C_{i,j}=I_{i,j}/I_{i,j}^{QS}$, where $C_{i,j}$, $I_{i,j}$, and $I_{i,j}^{QS}$ are the contrast, photometrically calibrated intensity, and photometrically calibrated intensity of QS for $i,j$ pixel on the image. 
Thus, we note that contrast values are dimentionless.

Next, the images were rotated to align the solar north pole at the top side. 
For this purpose, the data were separated into three groups. 
The first group comprises the data from BB, CT, Ka, Ko, La, MM, Mi, PM, and UP for which the alignment was achieved by compensating for ephemeris, i.e. rotating the image by the $p_0$ angle. 
However, for Ko it was necessary also to account for the rotation of the coelostat as was done by \cite{jha_long-term_2022}. 
The second group comprises most of the Ar, MD, and Co data which have been aligned prior to their digitisation. 
We note though that a 90$^\circ$ rotation was needed for the Ar data.
The last group comprises the Kh, Bo, and SP data which we aligned by using a cross-correlation approach. 
We used near-co-temporal observations from other archives that have been aligned as discussed previously to appropriately align the data from this third group. 
Data taken within a 7-day interval centred at the date of each observation were considered, but selecting the one closest in time. 
Correction for the differential rotation was applied, with SolarSoft routines \citep{freeland_data_1998}, so to project the reference image to the time of the image that was aimed to be aligned. 
In case no observation from any other archive was found within that time frame we used the closest in time observation from the same archive that has previously been aligned to act as the reference. 
The cross-correlation approach was also used to validate the orientation of the data in the other two groups too.
This was done 
to check whether the data in the second group were oriented correctly during the digitisation, but also because there might be errors in the date and time of observations which would result in a wrong orientation of the images (we remind that determination of the image rotation with ephemeris required accurate knowledge of the observations date and time).
Observations for which the resulting angle with the cross-correlation technique deviated significantly from the previously defined angle were tagged as potentially wrongly oriented, otherwise the previously defined angle was used. 
The main reason for such errors was mistakes in dating of the images, which thus allowed us to identify and correct wrong observational dates and times in the various archives.
We note, however, that such large errors sometimes occur also because of differences between the two observations (e.g. large scale artefacts not entirely accounted for or image distortions). It is noteworthy that most Ko and SP data have markings on the plate to denote the orientation, which could be used to identify the poles. 
However, our approach allowed for better versatility, while we also identified mistakes in the placement of the markings of the observations.

Figure \ref{fig:processedimages_flat} shows the observations shown in Fig. \ref{fig:processedimages_raw} after their calibration to produce contrast images and after the image rotation to show the north pole at the top.
Figure \ref{fig:processedimages_flat} demonstrates the effect on the observed solar features of several representative bandwidths used for acquisition of  H$\alpha$ data at various sites.
In particular, the contrast of the plage and filament regions is greater for the narrow band observations, e.g. MD and SP, while they are fainter for broader band observations, e.g. Ka and Ko. 
We notice that the contrast of features in Co data is lower than that in MD, even though the nominal bandpass is roughly the same. 
This suggests that Co data might have been taken slightly off-band or with a broader bandwidth compared to the MD ones.
We note, however, that also the transmission profile affects the contrast.
Furthermore, the information we have is about the nominal bandwidths, while the actual ones can differ due to instrumental setups, but also, especially for the spectroheliograms, due to potential adjustments or tests made by the observers.
This highlights the importance of having information on potential inconsistencies within the various archives.
Appendix \ref{sec:archive_characteristics} presents information about the characteristics of the archives that allow to assess their homogeneity in time, such as the evolution of disc eccentricity, large scale inhomogeneities, spatial resolution, and a comparison between the measured CLV in the images to the average from CT data.

Finally, on all the processed observations we identified filaments by using a modification of the multiple level tracking algorithm \citep{bovelet_new_2001}.
We applied two intensity thresholds (-3 and -5) as multiplicative factors to the standard deviation of the QS regions.
This gave two separate sets of masks where the pixels with contrast values above each threshold have the value of unity, while the value of 0 is assigned to all remaining pixels.
We then applied morphological operators to dilate and erode the mask obtained with the lower threshold.
These operators were defined to have widths in fractions of the solar disc radius to render them consistent for all archives.
We also applied a size threshold to the identified features to remove small-scale structures which are most likely artefacts. 
We kept only those features in the mask derived with the low contrast threshold that have at least one pixel in common with the features identified with the more conservative threshold.
This processing, however, still detects spots (which generally appear diminished in H$\alpha$ compared to continuum observations) as well as potential artefacts, such as scratches or hair introduced during the digitisation.
In order to remove such artefacts and not mistakenly treat them as filaments, we applied a linear fit and an ellipse fit to the mask resulting from each identified region.
This allowed us to exclude regions that are too circular, being most likely spots, or too straight and thin segments, in which case they would most likely be artefacts. 
For a rather small number of images this process results in oversegmentation and hence an overestimation of filament areas. 
We identified and excluded such cases with a conservative upper threshold on disc-integrated filament areas of 10\%. 
The pole markings were also removed from the SP data by removing all elongated vertical features found at the poles.
Examples of masks with identified filaments are shown in Fig. \ref{fig:processedimages_mask}.

\begin{sidewaysfigure*}
	\centering
		\includegraphics[width=1\linewidth]{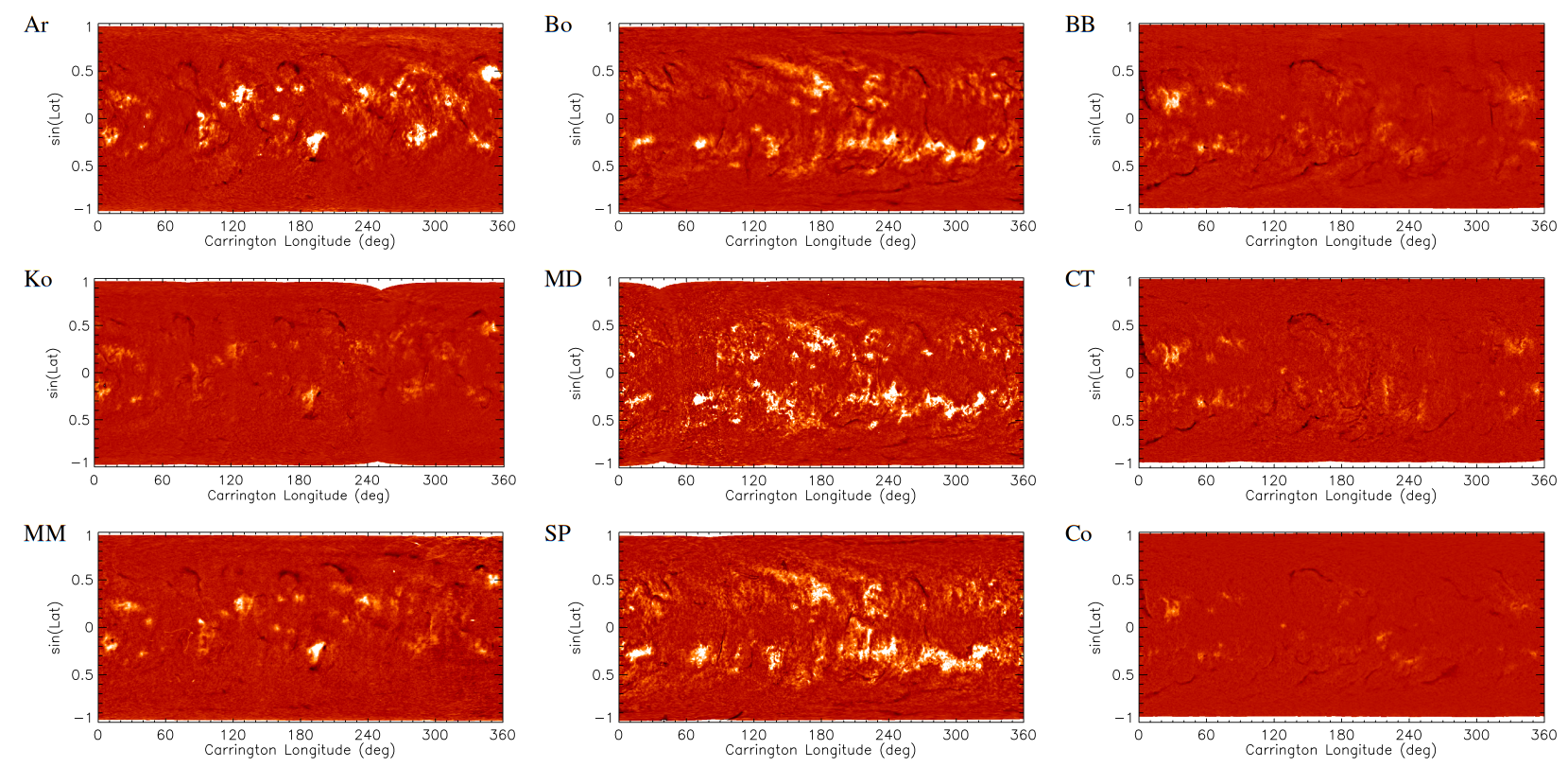}
\caption{Carrington maps constructed from the observations of nine H$\alpha$ archives analysed in this study. Each column shows Carrington maps for the same rotations: 1427 (7 May to 2 June 1960, left, see also left column of Fig. \ref{fig:comparisoncarrington2}), 1689 (30 November to 26 December 1979, middle), and 2141 (1 to 27 September 2013, right, see also bottom row of Figs. \ref{fig:processedimages_raw}--\ref{fig:processedimages_mask}). All maps show contrast values in the range [-0.5,0.5].}
	\label{fig:carringtonmaps}
\end{sidewaysfigure*}

\begin{figure*}
	\centering
		\includegraphics[width=1\linewidth]{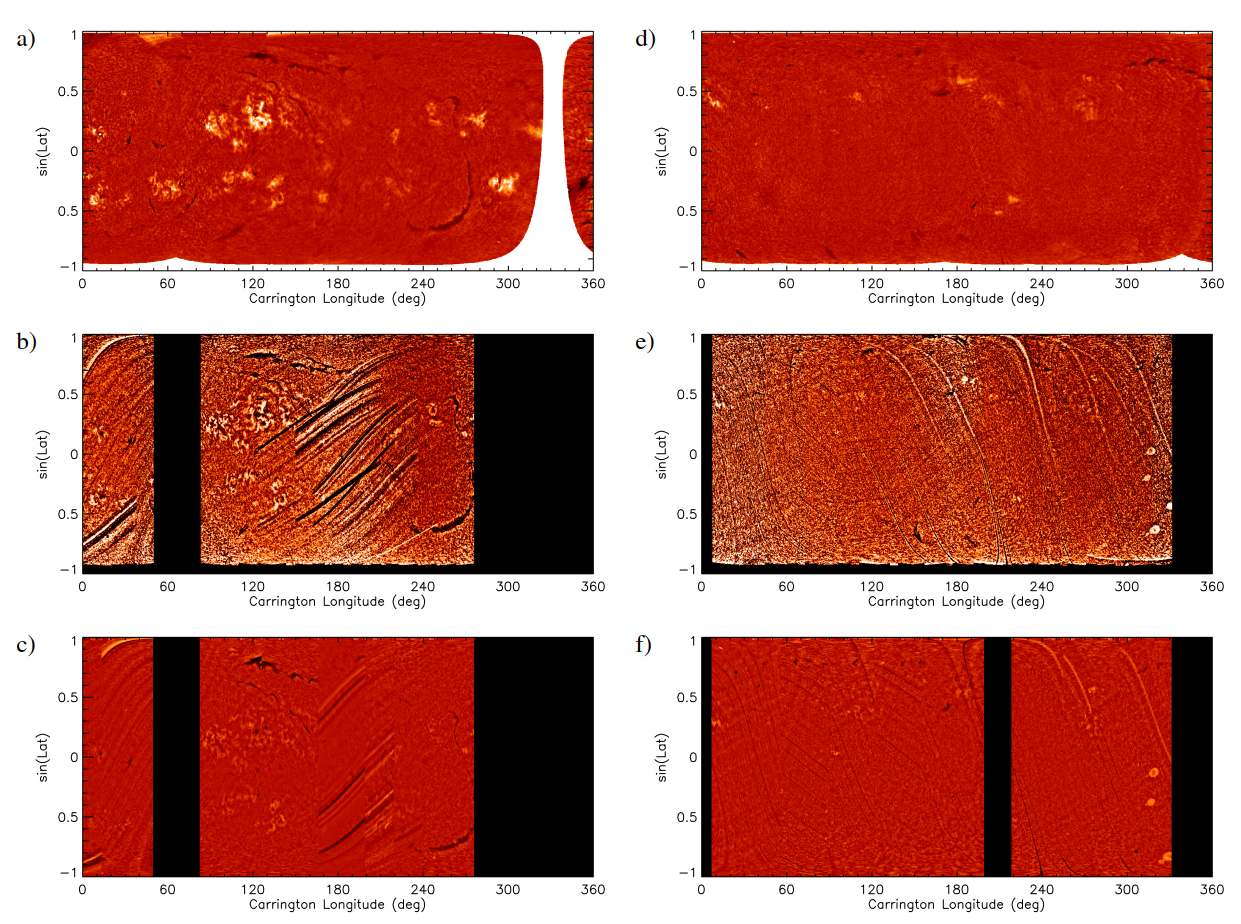}
		\caption{Comparison of Ko H$\alpha$ maps for Carrington rotations 854 (23 July to 18 August 1917, left) and 1497 (29 July to 25 August 1965, right, see also top row of Figs. \ref{fig:processedimages_raw}--\ref{fig:processedimages_mask} and right column of Fig. \ref{fig:comparisoncarrington2}) processed in this study (top row), by \citet[][middle row]{chatterjee_long-term_2017}, and by \citet[][bottom row]{lin_new_2020}. }
	\label{fig:comparisoncarrington1}
\end{figure*}

\begin{figure*}
	\centering
		\includegraphics[width=1\linewidth]{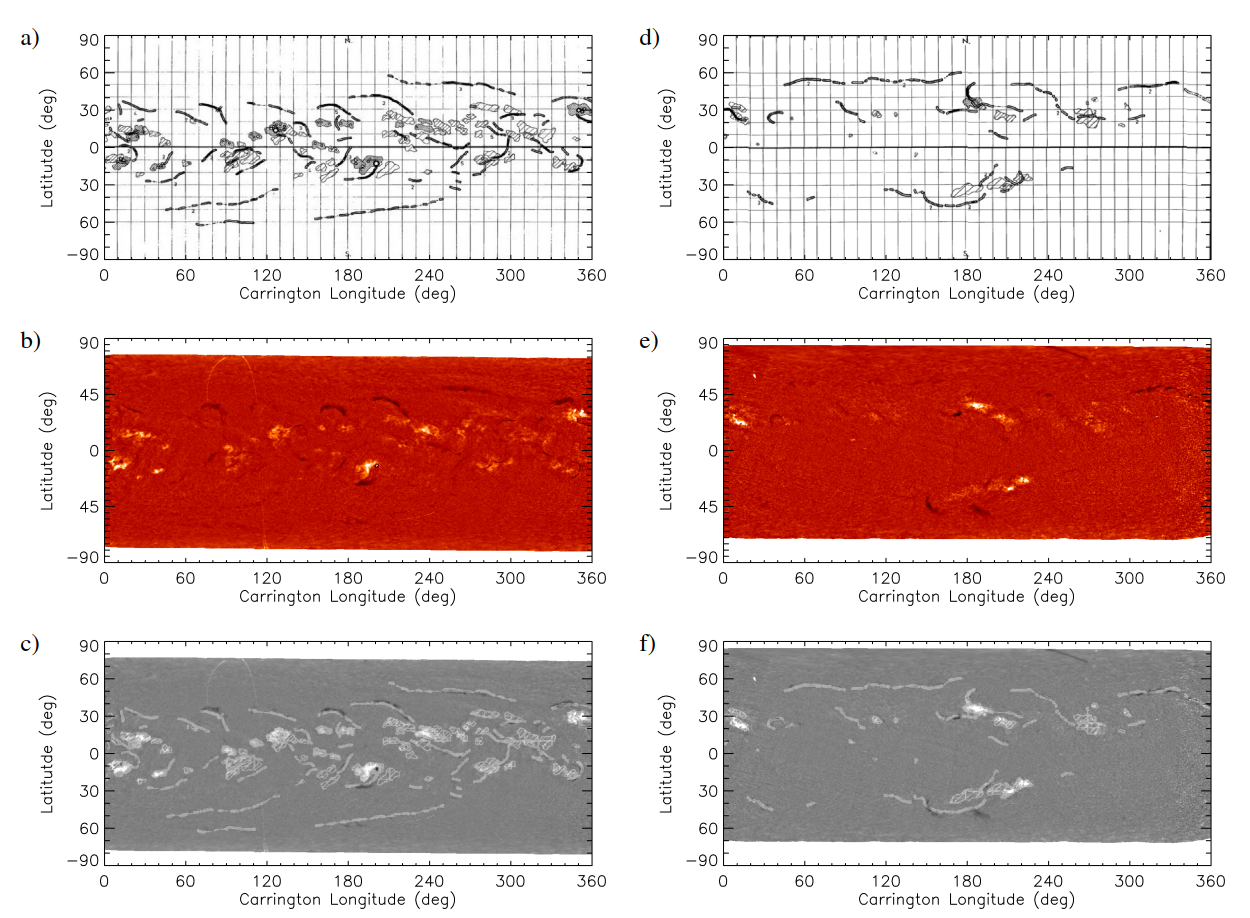}
		\caption{Comparison of MD H$\alpha$ maps for Carrington rotations 1427 (7 May to 2 June 1960, left, see also middle column of Fig. \ref{fig:carringtonmaps}) and 1497 (29 July to 25 August 1965, right, see also bottom row of Figs. \ref{fig:processedimages_raw}--\ref{fig:processedimages_mask} and right column of Fig. \ref{fig:comparisoncarrington1}) produced manually by observations at Meudon site (top row) and processed in this study (middle row). The bottom row shows the map obtained by overlapping the two corresponding maps shown in the upper two rows. To help visibility, this map is shown in grey-scale with the filaments from the manual processing (top panel) shown in light grey instead of black. }
	\label{fig:comparisoncarrington2}
\end{figure*}

\begin{figure*}
	\centering
		\includegraphics[width=0.82\linewidth]{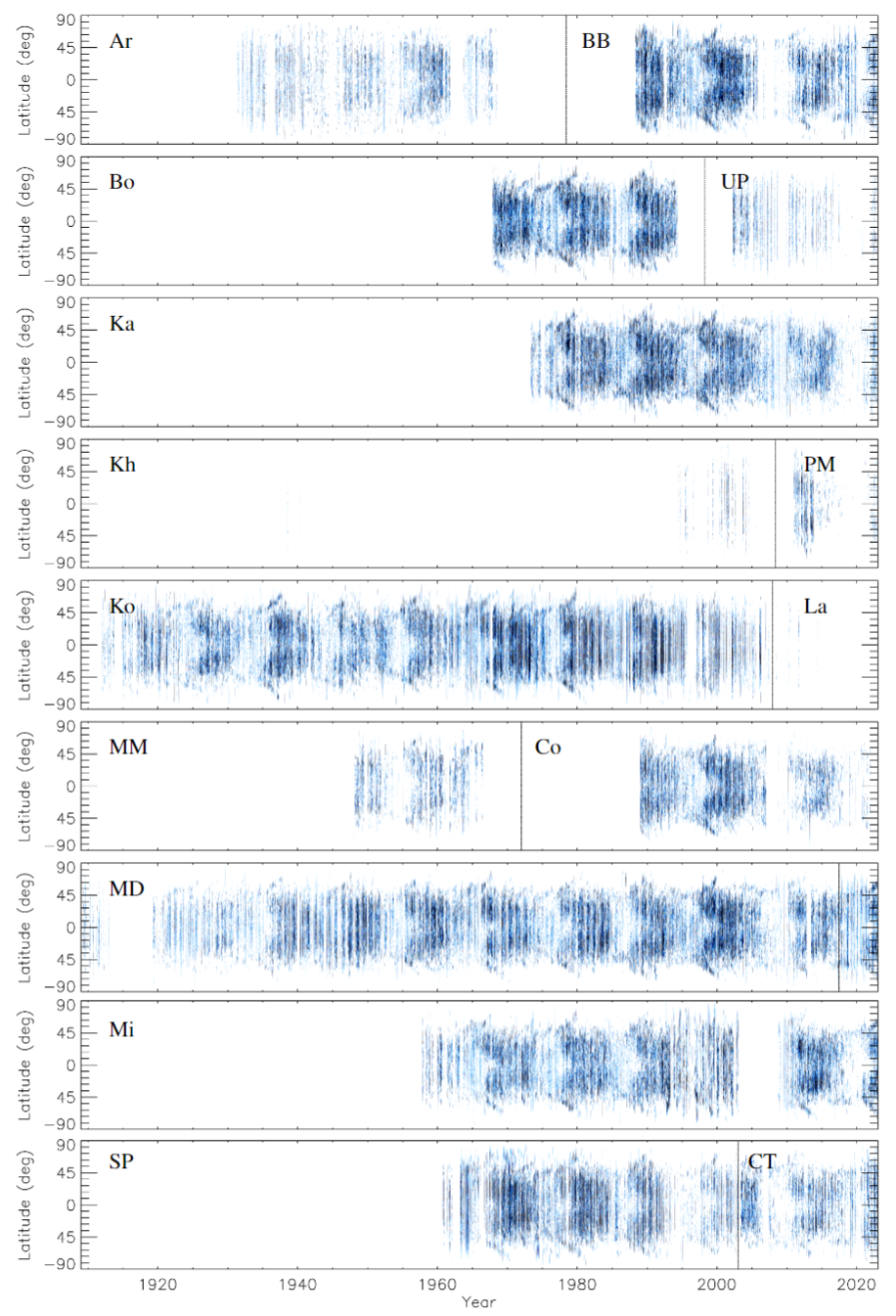}
		\caption{Filament butterfly diagrams constructed from the observations of all H$\alpha$ archives analysed in this study. The diagrams show daily mean areas within latitudinal strips of 1$^\circ$ as fractions of the area of the entire solar disc. 
White means no observed filaments, while blue denotes the filament fractional areas getting darker with growing areas
up to the saturation level of 2$\times10^{-4}$. 
A vertical black line marks the separation of archives in the case there are more than one included in a panel, except for MD where it marks the instrument upgrade over 2017. }
	\label{fig:butterfly1}
\end{figure*}

\begin{figure*}
	\centering
	\includegraphics[width=0.9\linewidth]{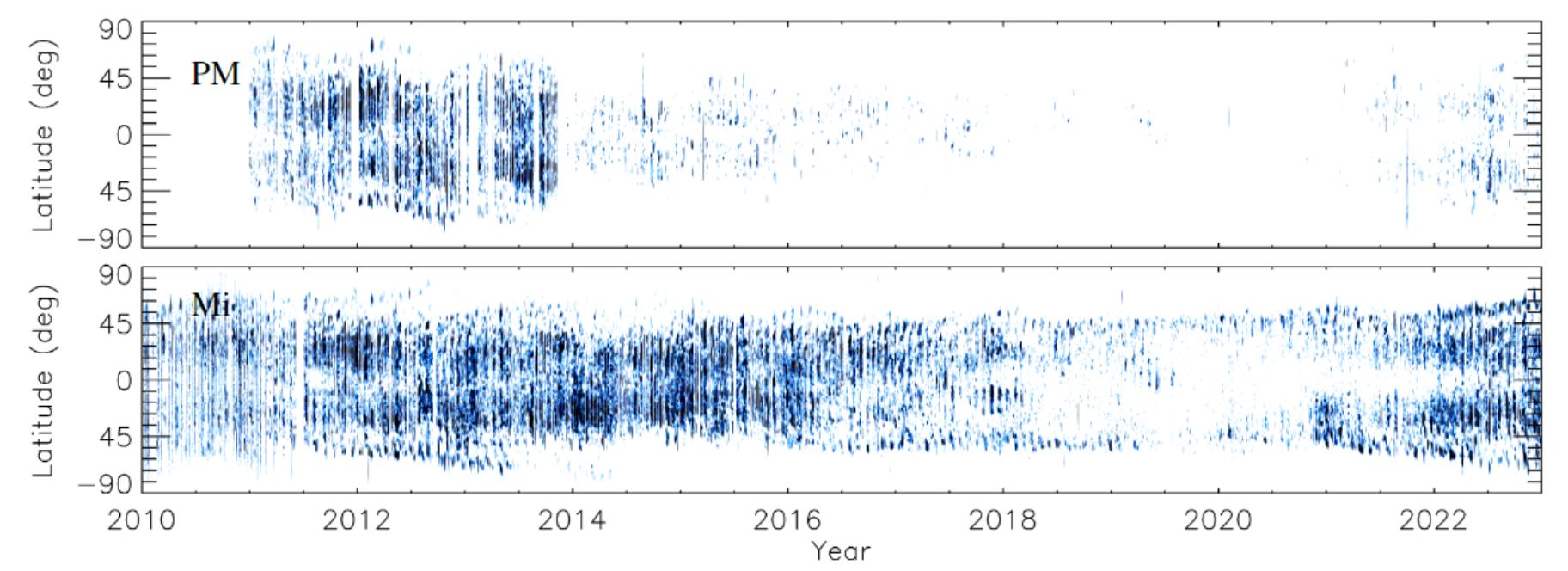}
		\caption{Filament butterfly diagrams constructed from the observations of PM (top) H$\alpha$ archive compared to those from Mi (bottom). The diagrams show daily mean filament areas within latitudinal strips of 1$^\circ$ as fractions of the area of the entire solar disc. White means no observed filaments, while blue denotes the filament fractional areas getting darker with growing areas
up to the saturation level of 2$\times10^{-4}$}
	\label{fig:butterfly2}
\end{figure*}

\begin{figure*}
	\centering
		\includegraphics[width=0.8\linewidth]{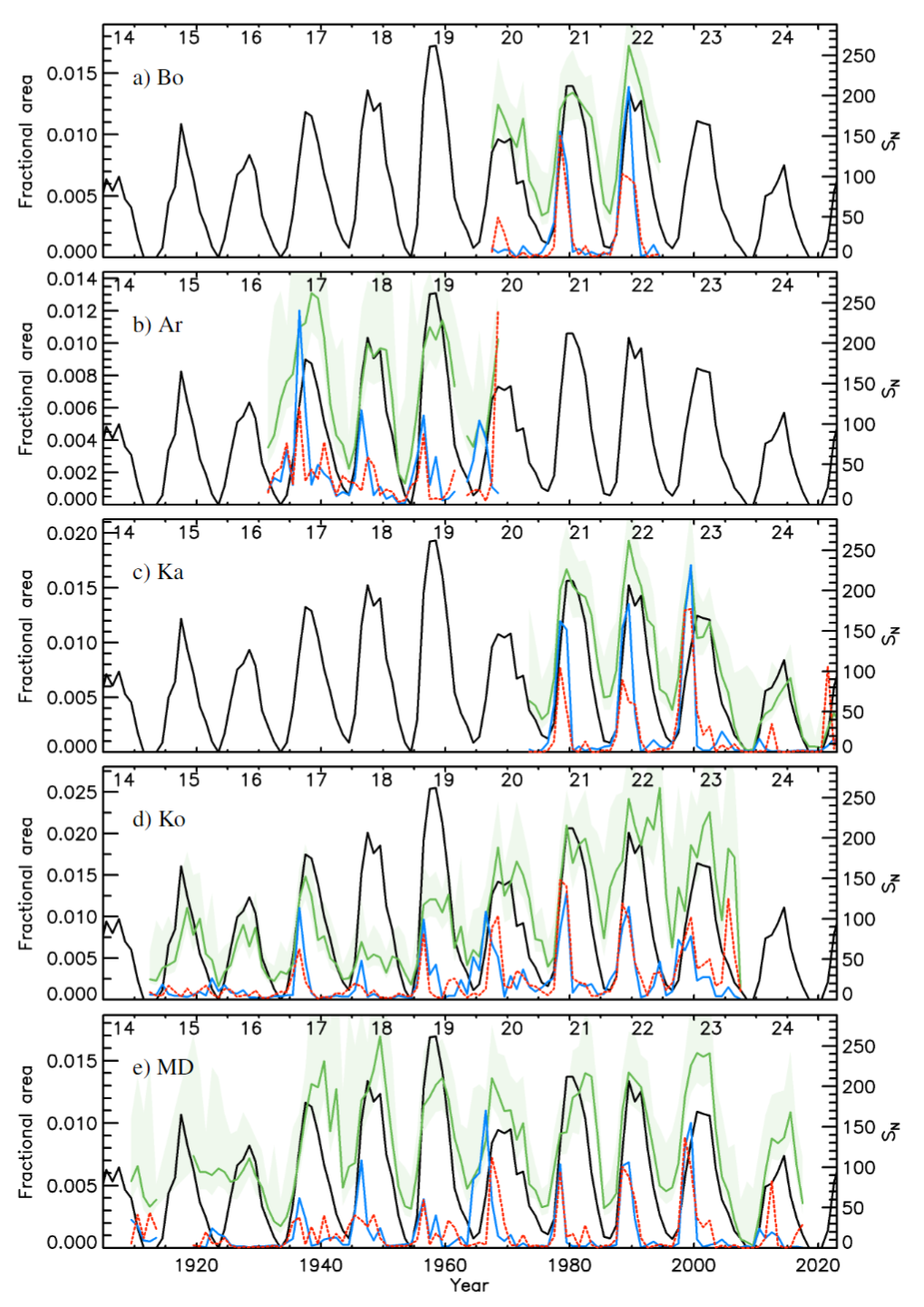}
		\caption{Filament fractional areas derived from the Bo (a), Ar (b), Ka (c), Ko (d), and MD (e) H$\alpha$ archives analysed in this study. Shown are annual median values of the disc integrated areas (green) as well as polar filament fractional areas (multiplied by 10) for latitudes -90$^\circ$ -- -50$^\circ$ (red) and 50$^\circ$ -- 90$^\circ$ (blue). The green shaded surface denotes the asymmetric 1-$\sigma$ interval of the disc-integrated areas. Also shown in black is the ISN series with its values given in the right $y$ axis. The numbers within the panels denote the conventional solar cycle numbering.}
	\label{fig:filamentsfractionalarea_individual}
\end{figure*}

\begin{figure*}
	\centering
	\begin{minipage}{0.90\linewidth}
\begin{overpic}[width=1\linewidth,trim={0 0cm 0cm 0.0cm},clip]{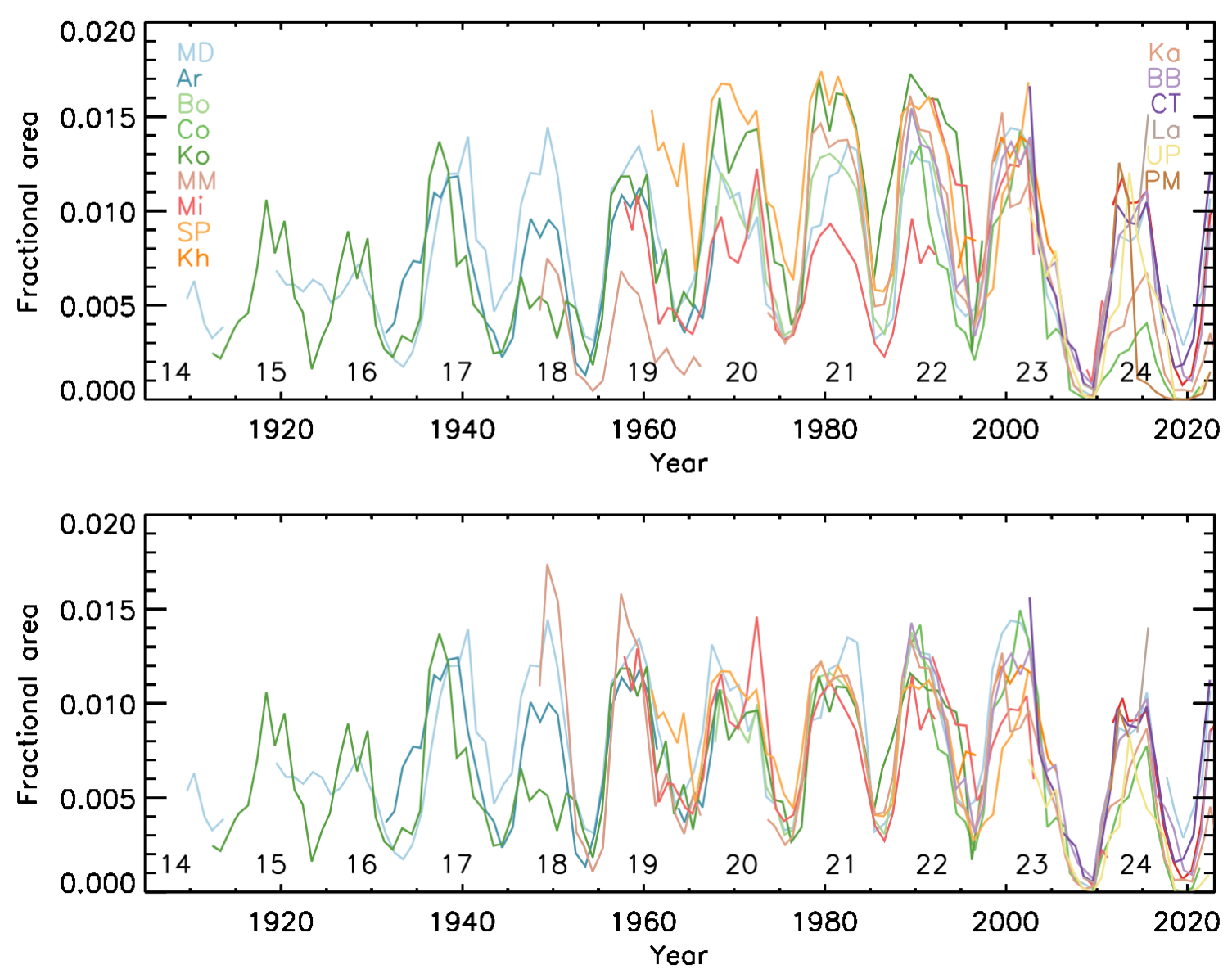}
	\put (13.00000,11.0000) {}     \end{overpic}
\end{minipage}
		\caption{Filament fractional areas from all analysed archives. Top panel shows the original areas, while the lower panel shows the areas after linearly scaling them to match those from MD. The numbers within the panels denote the conventional solar cycle numbering.}
	\label{fig:filamentsfractionalarea2}
\end{figure*}

\begin{figure*}
	\centering
	\begin{minipage}{0.90\linewidth}
\begin{overpic}[width=1\linewidth,trim={0 0cm 0cm 0.0cm},clip]{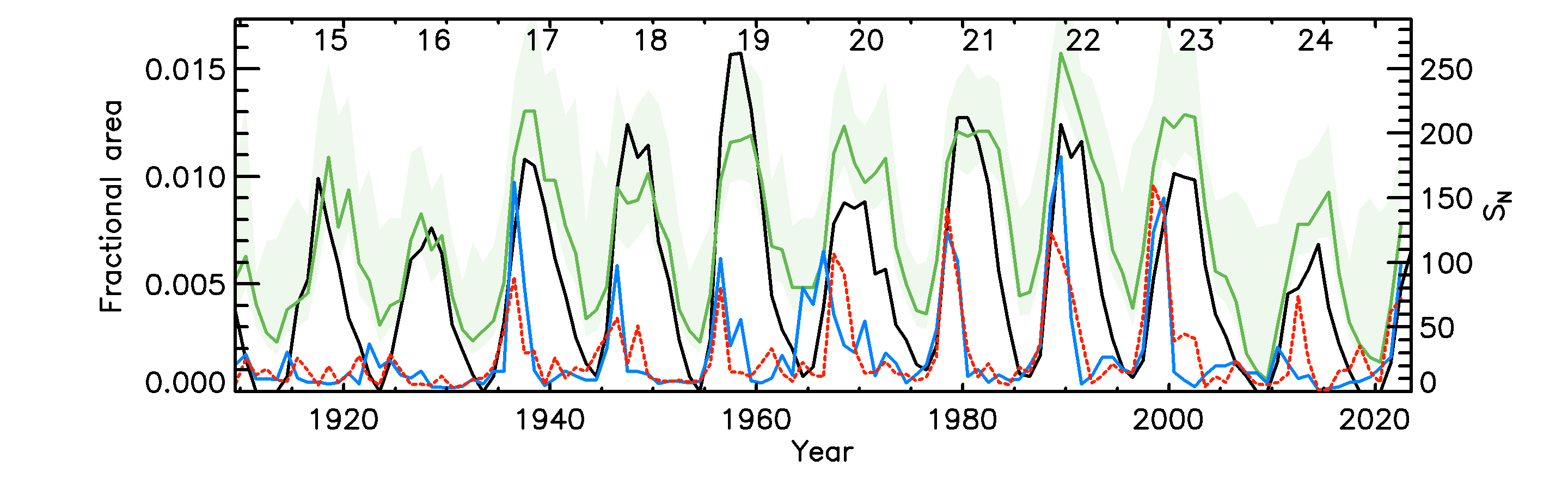}
	\put (13.00000,11.0000) {}     \end{overpic}
\end{minipage}
		\caption{		Filament fractional areas derived from the composite filament area series. Shown are annual median values of the disc integrated areas (green) as well as polar filament fractional areas (multiplied by 10) for latitudes -90$^\circ$ -- -50$^\circ$ (red) and 50$^\circ$ -- 90$^\circ$ (blue). The green shaded surface denotes the asymmetric 1-$\sigma$ interval of the disc-integrated areas. Also shown in black is the ISN series with its values given in the right $y$ axis. The numbers within the panels denote the conventional solar cycle numbering.}
	\label{fig:filamentsfractionalarea}
\end{figure*}

\begin{figure*}
	\centering
	\begin{minipage}{0.90\linewidth}
\begin{overpic}[width=1\linewidth,trim={0 0cm 0cm 0.0cm},clip]{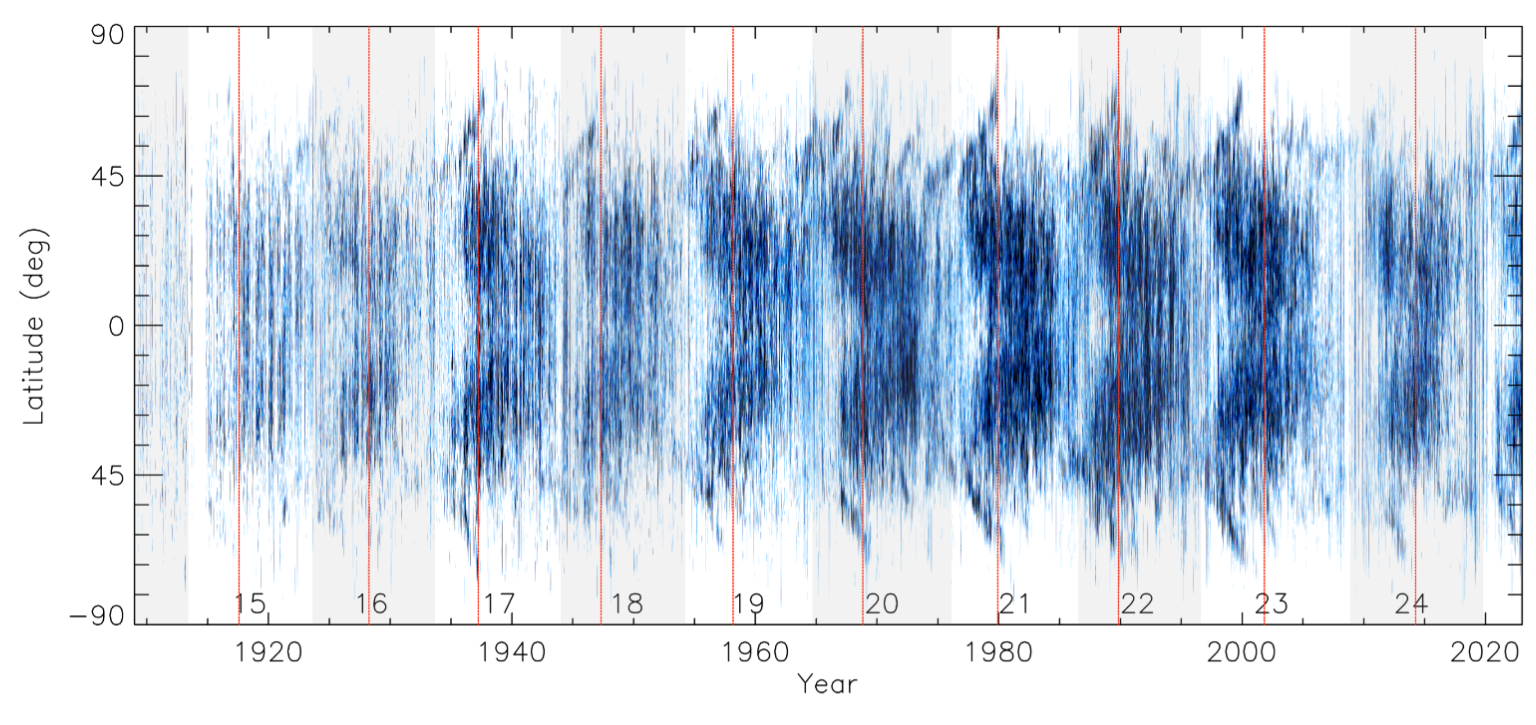}
	\put (13.00000,11.0000) {}     \end{overpic}
\end{minipage}
		\caption{Composite filament butterfly diagram constructed from the observations of all H$\alpha$ archives analysed in this study, except for PM after 2014 and Ko after 1998 (see Sects. \ref{sec:results} and \ref{sec:data}). The diagram shows daily mean areas within latitudinal strips of 1$^\circ$ as fractions of the area of the entire solar disc. White means no observed filaments, while blue denotes the filament fractional areas getting darker with growing areas
up to the saturation level of 3$\times10^{-4}$.  
		Even-numbered cycles are shaded in grey. The numbers at the lower part of the panel denote the conventional solar cycle numbering, while the date of the maximum of each cycle, as defined by Sunspot Index and Long-term Solar Observations (SILSO), is marked with a vertical red line.}
	\label{fig:butterfly3}
\end{figure*}

\section{Results}
\label{sec:results}

Using the 15 H$\alpha$ datasets (see Table~\ref{tab:observatories}) processed as described in the previous section, we have produced Carrington maps and butterfly diagrams of filament areas. In the following we present and discuss our results.

\subsection{Carrington maps}
Carrington maps are Mercator projections \citep{Calabretta_representations_2002} of the entire solar surface over one solar rotation.
We produced Carrington maps from all archives by considering only the regions within $\pm50^\circ$ longitudes of each image. 
However, this approach can sometimes lead to gaps in the produced Carrington maps. 
We fill such gaps, to the degree possible, with the values from another Carrington map which is produced by considering the entire solar disc from each image.
Overlapping images are averaged so to produce the final maps.

Figure \ref{fig:carringtonmaps} shows examples of produced Carrington maps from nine archives analysed in this study over three Carrington rotations.
Comparison of images in Fig. \ref{fig:carringtonmaps}  highlights the differences among the various archives.
In particular, the Carrington map from Ar data exhibits higher contrast than those from Ko and MM of the same rotation.
Similarly, filaments are more pronounced in Bo and CT Carrington maps compared to those for the same rotation from MD and Co, respectively. 
This might suggest that the observations taken with a very narrow bandwidth, like those at MD and Co, might not be the optimum ones if one wants to study the evolution of filaments.
However, we stress that archives such as Co and MD still allow performing quantitative studies of filaments.

The accuracy of the processing applied in our study is demonstrated by Figure \ref{fig:comparisoncarrington1}, which compares our Carrington maps based on Ko data to those by \cite{chatterjee_long-term_2017}\footnote{The Carrington maps are available at \url{https://kso.iiap.res.in/new/H_alpha}} and \cite{lin_new_2020}\footnote{Available at \url{https://sun.bao.ac.cn/hsos_data/GHA/synMap/KODA/}}.
As representative examples, we show the Carrington maps for rotations 854 (23 July to 18 August 1917) and 1497 (29 July to 25 August 1965).
The images over these intervals suffer from many artefacts, mostly dark linear segments close to the limb. 
These artefacts are clearly visible in maps produced by \cite{chatterjee_long-term_2017} and \cite{lin_new_2020}, while most of them have been accounted for by our method. 
We notice that the data artefacts lead to filaments being missed in the maps by \cite{chatterjee_long-term_2017} and \cite{lin_new_2020}, such as the one for rotation 1497 for longitudes 300--360$^\circ$ in the north hemisphere. 
However, we also notice residual artefacts in our produced maps, which are absent in the maps by \cite{chatterjee_long-term_2017} and \cite{lin_new_2020}. 
For example, the darker artefact for rotation 1497 for longitude 30$^\circ$ in the southern hemisphere. 
These are mostly small and clear artefacts close to the limb, which are magnified with their projection in this grid. 
These regions are removed in the maps by \cite{chatterjee_long-term_2017} and \cite{lin_new_2020} since they consider a smaller part of the disc when they construct the Carrington maps. 
Because of data gaps, the approach of those authors, however, results in big gaps in the produced maps, which is why we decided to keep a slightly larger region.
We also notice that the contrast of plage regions is diminished in the maps by \cite{chatterjee_long-term_2017} and \cite{lin_new_2020} compared to the ones derived from our processing, suggesting that their processing approach overestimates the background around plage regions \citep{chatzistergos_analysis_2018,chatzistergos_delving_2019,chatzistergos_full-disc_2022}, which might affect the contrast of filaments, too.
However, the lack of photometric  calibration might have also contributed to the decrease in contrast of features in the images with the processing by \cite{chatterjee_long-term_2017} and \cite{lin_new_2020}.

In Figure \ref{fig:comparisoncarrington2} we compare our Carrington maps produced with MD data to drawings from the same site produced manually by the observers \citep{laurenceau_retrospective_2015}\footnote{Available at \url{https://bass2000.obspm.fr/lastsynmap.php}} where they mark the locations of filaments and plage. 
We find a general agreement between the filaments seen in our calibrated images and the drawings. 
However, we also notice some disagreements, such as for rotation 1497 over longitudes 120--200$^\circ$ in the south hemisphere, for which the appearance of filaments is quite different in the two maps. 
We also notice that low contrast and fragmented filaments in our produced maps, such as the one for rotation 1497 over longitudes 40--180$^\circ$ in the north hemisphere, are reported as connected in the Meudon drawings.

We emphasise that any analysis of a single database suffers from the limited time coverage, as it may not capture the dynamic nature of filaments.
This limitation can however be alleviated by applying accurate and homogeneous processing to data from different archives, using as many data from different sources as possible, as done in our study.

\subsection{Filament areas}
Here we discuss and compare the resulting filament regions identified in the observations from the various H$\alpha$ datasets. 
We did the identification of filaments on individual images and not on the Carrington maps (as done in previous studies like \citealt{chatterjee_long-term_2017}), as this allows a more accurate identification of features. 
We then transformed the produced masks to a Mercator representation of the identified filaments.  
Figure \ref{fig:processedimages_mask} shows example masks of the identified filaments from observations of various datasets.
The identified features appear rather similar among data taken on the same day.
However, we also notice some differences, mainly due to the bandwidth or central wavelength used for the observations, which seems to influence the appearance of the filaments.
Some artefacts were also unfortunately still identified as filaments, for example the small and dark regions at the southern hemisphere of the Ar observations shown in Fig.  \ref{fig:processedimages_mask}.

Using the computed filament masks, we produced butterfly diagrams as time-versus-latitude maps. 
Figure \ref{fig:butterfly1} shows butterfly diagrams of the filament fractional area in bins of latitude of 1$^\circ$ from all individual archives. 
The poleward migration of filaments is clearly seen. 
This is evident for archives with good temporal coverage (e.g., BB, Bo, Ka, Ko, MD, Mi, SP), and less clear for archives with sparse observations (e.g., Ar, Kh, La, UP).
We also notice differences in the produced butterfly diagrams, largely due to the differences of the archives.
We do not notice, however, any clear change in the butterfly diagram due to the transition from plates to CCD, as in the cases of Co, MD, Ka, or Mi.
An interesting case is that of PM for which the data appear to be taken off-band since late 2013. 
The resulting butterfly diagram, shown also in Fig. \ref{fig:butterfly2} enlarged compared to Fig. \ref{fig:butterfly1}, displays the transition very clearly, with very few to no filaments registered after 2013.
This suggests that PM data after 2013 are, unfortunately, not a good source of information on filaments.

Figure \ref{fig:filamentsfractionalarea_individual} shows the disc integrated filament areas from five long running historical datasets analysed in this study (Bo, Ar, Ka, Ko, and MD) along with the international sunspot number \citep[ISN, hereafter;][]{clette_recalibration_2023} series.
The figure also includes the asymmetric 1-$\sigma$ intervals (determined separately for the values that are higher and lower than the annual mean values) as a measure of uncertainty and spread of values within a year.
We notice a different behaviour among the various archives.
For instance, the disc integrated filament areas increase with time in the Ko series, which parallels the behaviour of plage areas in Ko \ca data \citep{chatzistergos_analysis_2019,chatzistergos_delving_2019} suggesting that the increasing trend of quantities shown by both series is at least partly affected by the varying quality of Ko data (see also Appendix \ref{sec:archive_characteristics}).
In general, we do not see the disc integrated filament areas to exhibit the same cycle ranking as ISN.
Particularly, cycle 19, which is the strongest cycle in ISN, is comparatively ``normal'' in filament areas in all archives we analysed in this study.
We note that over the covered period there is agreement between ISN and the various group sunspot number series \citep[e.g.][]{svalgaard_reconstruction_2016,chatzistergos_new_2017,usoskin_robustness_2021}, and thus our discussion would not be affected by considering a different sunspot series.

In Fig. \ref{fig:filamentsfractionalarea_individual} we also show the areas of polar filament regions (multiplied by 10), defined as those at latitudes greater than $|50|^\circ$.
Polar filaments appear at the beginning of a solar cycle, and when the cycle reaches its maximum their areas decrease almost to zero.
However, we notice that the appearance of polar filaments for cycle 20 started earlier compared to other cycles.
The different series generally agree on the hemispheric asymmetry of polar filaments, although not always.
Notable exception is cycle 22 where MD and Ko show no asymmetry, while Bo and Ka show higher areas of polar filaments in the northern hemisphere.

Figure \ref{fig:filamentsfractionalarea2} shows annual mean disc integrated filament areas from all analysed archives. 
There are significant differences between the various series, but we remind that this is expected due to the different sampling in the different series and the observational differences of the various series \citep{chatzistergos_analysis_2020,chatzistergos_full-disc_2022}.
We remind that the appearance of filaments changes depending on the employed bandwidth (as well as filter transmission shape) of the observations, the central wavelength, but also the spatial resolution and thus the ability to resolve small filaments. Furthermore, since filaments are rather dynamic features significant changes might occur even within relatively short time differences between the observations. 
In order to check the consistency of the series, we linearly scaled the filament areas from all series to match those by MD.
The resulting filament areas are shown in the lower panel of Fig. \ref{fig:filamentsfractionalarea2} displaying an improved agreement after the scaling. 
However some discrepancies persist, such as over cycle 18 over which the filament areas from Ko are significantly lower than from the other series.

Finally, we combined results from all datasets together to produce an average series of filament fractional areas, which is shown in Fig. \ref{fig:filamentsfractionalarea}. 
For this we excluded PM data after 2014 as well as Ko data after 1998. 
Figure \ref{fig:filamentsfractionalarea} also shows the evolution of the disc integrated filament fractional areas in comparison to ISN. 
We find the filament fractional areas to be higher over cycles 20--23 than at earlier cycles or cycle 24, in contrast to ISN. 
We note that \cite{tlatov_tilt_2016-1} had reported a similar increase in number of filaments in MD data over cycles 22 and 23 compared to earlier cycles.
Figure \ref{fig:filamentsfractionalarea} includes also the average series of filament fractional areas over two latitudinal bands of [-90$^\circ$ -- -50$^\circ$] and [50$^\circ$ -- 90$^\circ$]. 
We find the rise of most cycles to largely coincide with that in ISN. 
Notable exceptions are cycle 20, over which we notice an earlier increase only in the north hemisphere, and cycle 23 for which the filament areas increase before ISN.
Our results also suggest very few polar filaments over cycles 15 and 16.

Figure \ref{fig:butterfly3} shows a composite butterfly diagram derived by combining the filament areas from all data analysed in this study after linearly scaling them to match those from MD.
We average all existing values on days where multiple archives have data.
This is, to our knowledge, the most complete filament butterfly diagram to-date.
The composite diagram brings forth the characteristics of filament evolution, in particular making more evident their poleward migration.

\section{Summary and conclusions}
\label{sec:conclusions}
Regular H$\alpha$ photographic observations of the solar disc date back to 1909. 
Several series of those observations have recently been made available in digital form.  

Observations in the H$\alpha$ line are a great resource of information on the chromosphere and particularly on filament regions. 
These fascinating solar features trace the magnetic polarity inversion lines and thus analysis of filament characteristics is important for understanding solar magnetism.
Filament data have the potential of aiding reconstructions of past solar magnetograms based on \ca observations \citep{chatzistergos_recovering_2019}, where H$\alpha$ observations can provide information on the polarity of regions.
\citet{mordvinov_long-term_2020} developed a method for the reconstruction of the solar magnetic field using the synoptic observations of the Sun's emission in the \ca and H$\alpha$ lines from Ko. 
Based on these reconstructed magnetic maps, they studied the evolution of the magnetic field in Cycles 15--19. 
Despite numerous past works with such data, a consistent and accurate processing of the various archives, including in particular the photometric calibration of historical photographic archives,  has been missing.

We analysed the most prominent historical archives currently available in digital form along with some modern CCD-based ones. 
The historical archives include those from Kodaikanal, Meudon, Kanzelhöhe, and Sacramento Peak which have been analysed before, but also those from Arcetri, Boulder, Coimbra, McMath-Hulbert, and Mitaka which, to our knowledge, have not been previously used to study filaments in an automatic way.
We recall that 9 of the 15 analysed series were unexplored until the present study, thus assessing their characteristics also allows establishing their potential for long-term studies of solar magnetism.
We showed that our processing accounted for most of the artefacts affecting the images better than was previous analyses of such data allowed.
We produced Carrington maps from the calibrated H$\alpha$ observations.
We also developed an automatic approach of identifying filaments in the images from the various archives. 
We produced butterfly (time/latitude) diagrams from each archive separately, as well as a composite diagram by considering all observations together, where we average all available observations on days when data from more than one observatory exist.
We note that the observational characteristics of the various archives differ significantly. This leads to sampling different heights of the solar atmosphere and thus affects the characteristics of the filament regions.
A more comprehensive study of the characteristics of the identified filaments from the different archives will be very important to help further understand the differences between the various archives.

It is worth noting that some of the archives have near-co-temporal observations in the \ca line, thus analysing these series can also provide insights into \ca data too.
We studied image characteristics of the various series finding common aspects with those previously reported for their \ca counterparts.
In particular, we find both \ca and H$\alpha$ observations from Ko to show a steady deterioration, or that both SP \ca and H$\alpha$ observations exhibit high disc ellipticity over the first $\sim$15 years. 
Finally, there are more historical H$\alpha$ archives that are not available in digital format at the moment, such as those from Baikal \citep{golovko_data_2002}, Crimea, Ebro \citep{curto_historical_2016}, Hamburg, Schauinsland, Wendelstein \citep{wohl_old_2005}, and Kislovodsk \citep{tlatov_tilt_2016-1}.
It would be very beneficial if those archives could be digitised, along with the remaining data from Kharkiv, in order to complement the available information of H$\alpha$ data and thus raise their reliability and value.

\begin{acknowledgements}
	The authors thank the observers at the Arcetri, Big Bear, Boulder, Catania, Coimbra, Kanzelhöhe, Kharkiv, Kodaikanal, Larissa, McMath-Hulbert, Meudon, Mitaka, Pic du Midi, Sacramento Peak, and Upice sites. 
	We thank Hardi Peter and Sami K. Solanki for the fruitful discussions.
	We also thank Isabelle Buale for all her efforts to digitise the Meudon archive.
	We thank the anonymous referee for the constructive comments that helped improve this paper. 
	We acknowledge Paris Observatory for the use of spectroheliograms. SP Spectroheliograms were acquired at Evans facility at Sac Peak operated by NSO/AURA/NSF, while the images were scanned by Dr. A. Tlatov (Russia).
	This work utilizes data from the National Solar Observatory Integrated Synoptic Program, which is operated by the Association of Universities for Research in Astronomy, under a cooperative agreement with the National Science Foundation and with additional financial support from the National Oceanic and Atmospheric Administration, the National Aeronautics and Space Administration, and the United States Air Force. The GONG network of instruments is hosted by the Big Bear Solar Observatory, High Altitude Observatory, Learmonth Solar Observatory, Udaipur Solar Observatory, Instituto de Astrofísica de Canarias, and Cerro Tololo Interamerican Observatory.
	H$\alpha$ data were provided by the Kanzelhöhe Observatory, University of Graz, Austria.
	Larissa observatory acknowledges funding from the Municipality of Larissa, Thessaly, Greece.
	This work was supported by grants PRIN-INAF-2014 and PRIN/MIUR 2012P2HRCR "Il Sole attivo", COST Action ES1005 "TOSCA", FP7 SOLID. 
	This research has received funding from the European Union's Horizon 2020 research and innovation program under grant agreement No 824135 (SOLARNET). 
	T.C. thanks ISSI for supporting the International Team 474 “What Determines The Dynamo Effectivity Of Solar Active Regions?”.
	TB, RG, and NP, acknowledge financial support by Funda\c{c}\~{a}o para a Ci\^{e}ncia e a Tecnologia (FCT) through the research grants UIDB/04434/2020 and UIDP/04434/2020.
	This research has made use of NASA's Astrophysics Data System (ADS; \url{https://ui.adsabs.harvard. edu/}) Bibliographic Services.
\end{acknowledgements}

\bibliographystyle{aa}
\bibliography{_biblio01}   

\appendix
\section{Archive characteristics}

\label{sec:archive_characteristics}
\begin{figure*}
	\centering
		\includegraphics[width=1\linewidth]{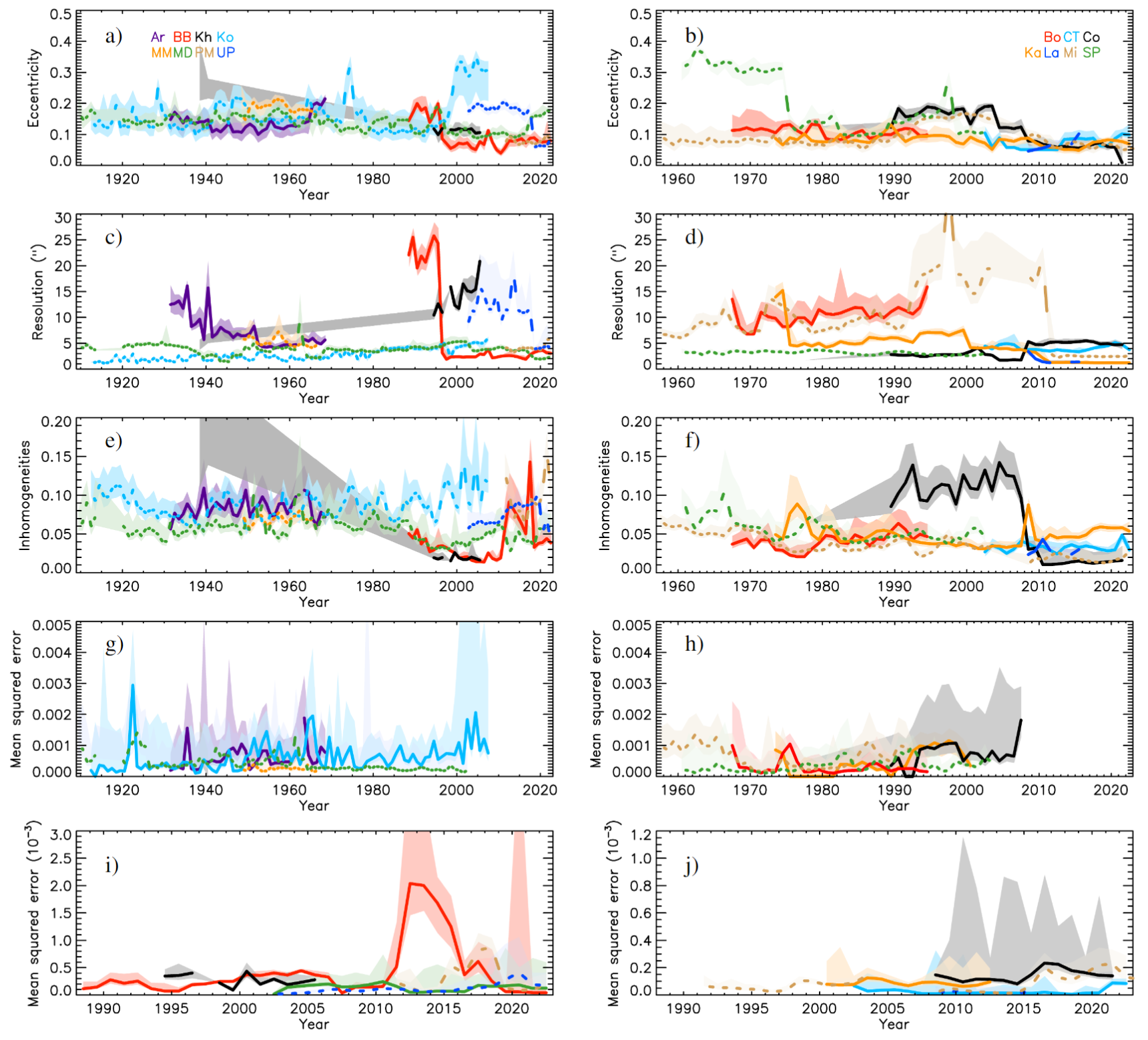}
\caption{Temporal evolution of key characteristics of the raw H$\alpha$ data analysed in this study: a) and b) eccentricity of the recorded solar disc; c) and d) spatial resolution; e) and f) large-scale inhomogeneities; and g)--j) the mean-squared error of the fit to the curve relating the measured QS CLV from the various archives analysed in this study to the reference QS CLV obtained from CT data. The latter is shown separately for the photographic data (panels g and h) and the CCD-based ones (panels i and j). 
The legend at the top panels give the color association of the different archives. 
Shown are annual median values, along with their asymmetric 1$\sigma$ intervals (shaded surfaces). See Appendix \ref{sec:archive_characteristics} for more details.}
	\label{fig:datasetstatistics}
\end{figure*}

Here we discuss some characteristics of the digital images from the various archives we analysed. 
Following \cite{ermolli_comparison_2009} and \citet{chatzistergos_delving_2019} we studied the solar disc eccentricity, the spatial resolution, the large scale inhomogeneities, and the mean standard error between the measured QS CLV to the reference QS CLV (taken to be the one from CT data). 
Table \ref{tab:datasetcharacteristics} and Fig. \ref{fig:datasetstatistics} show the results for the characteristics of the various archives used in this study.
Findings from this study helped us to optimize details in steps of our image processing, in addition to provide new information on the analysed archives. 
We recall that 9 of the 15 analysed series were unexplored until the present study, thus assessing their characteristics also allows establishing their potential for long-term studies of solar processes.

We first compared the solar disc eccentricity, defined as $e=\sqrt{1-(R_{\mathrm{min}}^2/R_{\mathrm{max}}^2)} $, where $R_{\mathrm{min}}$ and $R_{\mathrm{max}}$ are the semi-minor and -major axes as measured with a fit of an ellipse to the edge of the solar disc, which was identified with Sobel filtering \citep[for more details see Appendix B of][and \citealt{chatzistergos_historical_2020}]{chatzistergos_analysis_2019}.
We note that there are various reasons which can lead to the recorded solar disc having a roughly elliptical shape, including atmospheric diffraction \citep{corbard_importance_2019}, potential tilts between the optical axis and the camera, as well as issues during the digitisation of the plates. 
However, the most common reason for spectroheliograms to exhibit elliptical (and sometimes rather distorted) solar discs is due to an uneven motion of the instrument, which causes some rasters to be stretched unevenly compared to others in the recorded image. 
We find all archives to have rather similar eccentricities with mean values within 0.06 and 0.2.
The CCD-based data have slightly lower values.
We notice, however, that there are periods with increased disc ellipticity for some archives, for instance SP data before 1975, BB data before 1996, Co data before 2008, or Ko data over 1972--1976 and after 1998.
The results for SP are consistent to those for the SP \ca data for the increased ellipticity before 1975 \citep{chatzistergos_historical_2020}.
The decrease of ellipticity in the Co data coincides with the introduction of the CCD camera. 
We note that a similar decrease is seen in MD data over 2002 when the CCD camera was introduced, but the change is smaller compared to the one for Co data.
Concerning the Ko data after 1998 we note that the values for the ellipticity (and most metrics in the following) are not reliable because almost all of those data have large parts of the disc missing (see Sect. \ref{sec:data}).

We compared the spatial resolution of the observations by finding the frequency for which the 98\% of the power spectral density of the solar disc is taken into account. Following \cite{chatzistergos_delving_2019} we did the computation within 64$\times$64 pixel$^2$ sub-arrays of QS regions. This was repeated 100 times by randomly selecting a different QS region across the inner $R/3$ of the solar disc. The mean value of the 100 computations was adopted as the mean spatial resolution of the images.
Most archives analysed in this study exhibit a rather stable spatial resolution with time.
The spatial resolution for the Ko data is almost constantly increasing with time suggesting worsening conditions with time. This is consistent with the results for the Ko \ca data \citep{chatzistergos_delving_2019}.
The spatial resolution of the MD observations is rather stable with time. We notice a decrease in spatial resolution with time for Ar, Ka, and BB data, while Mi data increase after 1990 and decrease after 2011. For BB, Ka, and Mi data it is mostly due to changes in pixel scale of the observations. The decrease in spatial resolution of Ar data is partly due to some images over the early periods having saturated regions. This is consistent with the Ar \ca data too \citep[][]{chatzistergos_analysis_2019}.

Then we evaluated the large-scale inhomogeneities affecting the observations. Following \cite{chatzistergos_delving_2019} we define a measure of the inhomogeneities as the relative difference of the computed background of the images to the radially symmetric QS CLV.
Most archives analysed in this study show a rather stable level of inhomogeneities with time.
We find a decrease in the inhomogeneities with time for the SP and MD archives, while there is an abrupt decrease for the Co data over 2008 coinciding with the installation of the CCD camera.
In contrast to that there is a considerable increase in the BB data after 2010. Many BB images over that period suffer from artefacts, probably introduced during the calibration of the CCD, rendering the centre of the disc darker than the regions close to the limb (see Fig. \ref{fig:processedimages_raw}). We note, however, that such issues are treated by our processing accurately enough.

Finally, we computed the mean standard error of the fit between the measured QS CLV and a standard reference QS CLV.
The reference QS CLV was the one we used for the photometric calibration of the historical data and was the mean QS CLV from selected CT data unaffected by large scale artefacts.
We remind that CT observations were performed with a CCD camera. 
When computing the mean errors we consider the photographic and CCD-based data separately because their results are not comparable.
That is because the photographic data are given in values of density, while the CCD-based ones are already in intensity units.
However, to make the results from the various CCD-based data comparable, we normalise the measured QS CLV pattern so that its maximum value is always one (this is needed due to the different bit depth of the images from different archives).
Furthermore, the errors from all archives (both photographic and CCD-based ones) have been divided by the degrees of freedom of the fit in each case to render them comparable to each other (this is needed because of the different solar disc dimentions in the images from different archives).
The resulting mean errors for all photographic archives are relatively low in the range 0.0001--0.0020, while the standard deviation of the errors reaches values up to 0.005 when ignoring the Ko data after 1998.
We notice in most archives that there are periods for which the errors increase abruptly, but these tend to be for isolated years.
For the Ko data there is a slight increase of the error with time, similar to the one reported for the Ko \ca data \citep{chatzistergos_analysis_2018,chatzistergos_delving_2019}, favouring the argument that this is due to worsening atmospheric conditions or instrumental wearing.
However, the increase of the error for the H$\alpha$ data is smaller compared to the one reported for the \ca data \citep{chatzistergos_delving_2019}. 
This might be due to the wavelength dependent behaviour of seeing, being less severe for longer wavelengths, in this case H$\alpha$ compared to \ca \citep{boyd_wavelength_1978,Rimmele_solar_2011} and thus lending support in the suggestion that seeing conditions at Kodaikanal deteriorated with time.
The errors for Ko data after 1998 reach a value of 0.12.
The periods for which the errors for the Ko data exhibit abrupt changes do not match those from the Ko \ca data, thus lowering the probability that these particular ones are due to instrumental issues.
However, we note that this is not a robust test, since it merely measures how different the QS CLV of the various data is to the reference set, while it might also be partly affected by severe large scale inhomogeneities affecting the images.
The results for the CCD-based data show even smaller variations (though we stress that the absolute values for the CCD-based and photographic data are not comparable).
CT data were used to derive the reference QS CLV pattern and they show rather low variations that are less than 0.2$\times10^{-3}$.
We notice relatively low errors for archives with similar bandwidth to CT, while the errors are slightly higher for MD and Co which have a narrower bandwidth.
We also notice a sharp increase in the mean error for PM since 2014, which would be in line with the observation that PM data since that period were taken off-band.
There is also a sharp increase in the mean error for BB data after 2011.
This is due to artefacts of the images that caused the centre of the disc to be darker than the regions near the limb.

The above analysis allowed us to identify archive inconsistencies, but also to verify some that were previously found in contemporaneous \ca data.
Thus, this has also implications for studies with \ca data from the sites that have contemporaneous data to H$\alpha$ and importantly for studies reconstructing past solar magnetism \citep{chatzistergos_full-disc_2022} and solar irradiance \citep{chatzistergos_long-term_2023}.

\begin{table*}
	\caption{List of determined characteristics of H$\alpha$ archives analysed in this study.}          
	\label{tab:datasetcharacteristics}   
	\centering                                  
	\small
	\begin{tabular}{l*{5}{c}}          
		\hline\hline                      
Dataset&Eccentricity&Resolution&Inhomogeneities&\multicolumn{2}{c}{Mean error}\\
&&&&Photographic ($10^{-2}$)& CCD ($10^{-3}$)\\
\hline
Ar &  0.13$\pm$ 0.05 &  6.54$\pm$ 3.79 &  0.09$\pm$ 0.03 &  0.08$\pm$ 0.11&-\\
BB &  0.10$\pm$ 0.06 &  7.59$\pm$ 9.03 &  0.04$\pm$ 0.04 & -&  0.45$\pm$ 1.36\\
Bo &  0.11$\pm$ 0.04 & 10.74$\pm$ 3.63 &  0.04$\pm$ 0.02 &  0.05$\pm$ 0.07&-\\
CT &  0.07$\pm$ 0.03 &  4.38$\pm$ 1.00 &  0.03$\pm$ 0.01 & -&  0.03$\pm$ 0.08\\
Co &  0.12$\pm$ 0.06 &  3.82$\pm$ 1.40 &  0.07$\pm$ 0.06 &  0.12$\pm$ 0.16&  0.18$\pm$ 0.40\\
Ka &  0.08$\pm$ 0.04 &  4.00$\pm$ 2.84 &  0.05$\pm$ 0.02 &  0.06$\pm$ 0.07&  0.10$\pm$ 0.11\\
Kh &  0.15$\pm$ 0.09 & 11.93$\pm$ 5.25 &  0.06$\pm$ 0.07 &  0.08$\pm$ 0.10&  0.28$\pm$ 0.18\\
Ko &  0.16$\pm$ 0.09 &  2.57$\pm$ 1.08 &  0.09$\pm$ 0.03 &  0.09$\pm$ 0.17&-\\
La &  0.06$\pm$ 0.02 &  1.88$\pm$ 0.89 &  0.03$\pm$ 0.01 & -&  0.03$\pm$ 0.03\\
MM &  0.19$\pm$ 0.04 &  5.81$\pm$ 2.18 &  0.07$\pm$ 0.02 &  0.04$\pm$ 0.04&-\\
MD &  0.13$\pm$ 0.04 &  3.74$\pm$ 1.25 &  0.05$\pm$ 0.03 &  0.06$\pm$ 0.10&  0.18$\pm$ 0.25\\
Mi &  0.08$\pm$ 0.04 &  9.33$\pm$ 6.73 &  0.04$\pm$ 0.02 &  0.07$\pm$ 0.08&  0.10$\pm$ 0.14\\
PM &  0.08$\pm$ 0.02 &  4.26$\pm$ 0.99 &  0.10$\pm$ 0.04 & -&  0.35$\pm$ 0.34\\
SP &  0.20$\pm$ 0.10 &  3.26$\pm$ 0.66 &  0.06$\pm$ 0.03 &  0.05$\pm$ 0.06 &-\\
UP &  0.15$\pm$ 0.06 & 11.53$\pm$ 7.81 &  0.07$\pm$ 0.02 & -&  0.19$\pm$ 0.32\\ 
\hline
	\end{tabular}
	\tablefoot{}
\end{table*}

\end{document}